\documentclass[prd,showpacs,twocolumn,nopacs,nofootinbib,superscriptaddress]{revtex4}
\usepackage{mathrsfs}
\usepackage{yfonts}
\usepackage{tipa}
\usepackage{bm}
\usepackage{bbm}
\usepackage{amsmath}
\usepackage{amssymb}
\usepackage{amsthm}
\usepackage{amsfonts}
\usepackage{mathtools}
\usepackage{latexsym}
\usepackage{relsize}
\usepackage[greek,english]{babel}
\usepackage{booktabs}
\usepackage[iso-8859-7]{inputenc}
\usepackage{tikz}
\usepackage{array}
\usepackage{cancel}
\usepackage{times}

\newcommand{\lp}{\left(}
\newcommand{\rp}{\right)}
\newcommand{\lb}{\left[}
\newcommand{\rb}{\right]}

\newcommand{\bea}{\begin{eqnarray}}
\newcommand{\eea}{\end{eqnarray}}
\newcommand{\be}{\begin{equation}}
\newcommand{\ee}{\end{equation}}

\newcommand{\D}{{\cal D}}

\newcommand{\md}{\mathrm{d}}
\newcommand{\gra}{\nabla}
\newcommand{\p}{\partial}
\newcommand{\g}{\sqrt{-g}}

\begin{document}

\title{$f(Q,T)$ gravity}

\author{Yixin Xu}
\email{xuyx27@mail2.sysu.edu.cn}
\affiliation{School of Physics, Sun Yat-Sen University, Xingang Road, Guangzhou 510275, People's
Republic of China}
\author{Guangjie Li}
\email{li3582@purdue.edu}
\affiliation{Department of Physics and Astronomy, Purdue University, 525 Northwestern Avenue, West Lafayette, Indiana 47907-2036, USA}
\affiliation{School of Physics, Sun Yat-Sen University, Xingang Road, Guangzhou 510275, People's
Republic of China}
\author{Tiberiu Harko}
\email{tiberiu.harko@gmail.com}
\affiliation{Department of Physics, Babes-Bolyai University, Kogalniceanu Street,
Cluj-Napoca 400084, Romania}
\affiliation{School of Physics, Sun Yat-Sen University, Xingang Road, Guangzhou 510275, People's
Republic of China}
\author{Shi-Dong Liang}
\email{stslsd@mail.sysu.edu.cn}
\affiliation{School of Physics, Sun Yat-Sen University, Xingang Road, Guangzhou 510275, People's
Republic of China}
\affiliation{State Key Laboratory of Optoelectronic Material and Technology}
\affiliation{Guangdong Province Key Laboratory of Display Material and Technology, Guangzhou, People's Republic of China}

\begin{abstract}
We propose an extension of the symmetric teleparallel gravity, in which  the gravitational action $L$ is given by an arbitrary function $f$ of the nonmetricity $Q$ and of the trace of the matter energy-momentum tensor $T$, so that $L=f(Q,T)$. The field equations of the theory are obtained by varying the gravitational action with respect to both metric and connection. The covariant divergence of the field equations is obtained, with  the geometry-matter coupling leading to the nonconservation of the energy-momentum tensor.  We  investigate  the cosmological implications of the theory,  and we obtain the cosmological evolution equations for a flat, homogeneous and isotropic geometry, which generalize the Friedmann equations of general relativity. We consider several cosmological models by imposing some simple functional forms of the function $f(Q,T)$, corresponding to additive expressions of $f(Q,T)$ of the form $f(Q,T)=\alpha Q+\beta T$, $f(Q,T)=\alpha Q^{n+1}+\beta T$, and $f(Q,T)=-\alpha Q-\beta T^2$.  The Hubble function, the deceleration parameter, and the matter energy density are obtained as a function of the redshift by using analytical and numerical techniques. For all considered cases the Universe experiences an accelerating expansion, ending with a de Sitter type evolution. The theoretical predictions  are also compared with the results of the standard $\Lambda$CDM model.
\end{abstract}

\maketitle
\tableofcontents

\section{Introduction}

The development of gravitational theories closely followed the advances in differential geometry. In all geometric descriptions of gravity it is assumed, following \cite{Riemm}, that the space-time is endowed with a metric structure in a general space based on the element of arc $ds=F\left(x^1,...,x^n; dx^1,...,dx^n\right)$, where $F(x; y)$ is a positive (for $y \neq 0$) function defined  on the tangent bundle $TM$. Moreover, it is generally assumed that $F$ is homogeneous of degree one in $y$ \cite{Chern}. An important special case is represented by the choice $F^2=g_{\mu \nu}dx^{\mu}dx^{\nu}$, with the corresponding geometry called generally Riemannian geometry. Riemannian geometry lays at the foundations of general relativity \cite{Ein1, Hilbert, Ein2}, a geometric theory of gravity, which has become, together with quantum mechanics,  one of the cornerstones of present day physics.  General relativity is presently considered to be the most successful theory of gravity ever proposed. Its remarkable predictions on the perihelion advance of Mercury, on the deflection of light by the Sun, gravitational redshift \cite{Landau}, or radar echo delay \cite{Shap1,Shap2} have been confirmed observationally at an unparalleled level of accuracy. Moreover,  predictions such as the orbital decay of the Hulse-Taylor binary pulsar, due to gravitational - wave damping, have also fully confirmed the observationally  weak-field validity of the theory \cite{Taylor}. For a review on the experimental and observational tests of general relativity  see \cite{Will}. The detection of the gravitational waves \cite{gravwave}  did give the opportunity to evaluate the predictions of general relativity  in the final stages of binary black hole coalescence, corresponding to the limiting case of the strong gravitational fields.

On the other hand recent observational advances in cosmology have provided strong evidence that recently our Universe did enter in an accelerated expansion phase \cite{1a,1b,1c,1d,1e,1f,1g, 1h}. Moreover, the same observations indicate the surprising result that around 95 - 96\% of the content of the Universe is in the form of two mysterious components, called dark energy and dark matter, respectively, with only about 4 - 5\% of the total composition represented by baryonic matter \cite{2a,2b}. These observations  have shown the limitations of standard general relativity, which despite its important achievements, and its  remarkable success at the Solar System scale, may not be adequate to fully explain gravitational phenomena on  galactic and cosmological ranges. Hence  standard general relativity  may not be the ultimate theory of the gravitational force, since it cannot give satisfactory explanations to the two fundamental problems present day cosmology is confronted with: the dark matter problem and the dark energy problem, respectively. Moreover, since Einstein's standard theory predicts the existence of space-time singularities in the Big Bang and  inside black holes,  general relativity represents an incomplete physical model. To solve the singularity problem a consistent prolongation of general relativity into the quantum domain is probably needed.

To explain the observational results of cosmology many different approaches at the classical level have been proposed recently. However,  a satisfactory theory of gravity has yet to be found. One possibility to build new gravitational theories is to assume that at large scales the Einstein gravity model of general relativity breaks down, and a more general action than the standard Hilbert one, given by $S=\int{\left(R/2\kappa ^2+L_m\right)\sqrt{-g}d^4x}$, where $R$ is the Ricci scalar, $\kappa $ is the gravitational coupling constant, and $\sqrt{-g}$ is the determinant of the metric tensor, respectively,  describes the gravitational field. An important theoretical direction of study is represented by approaches  in which, by keeping the geometrical background as strictly Riemannian,  the standard Hilbert-Einstein  action is replaced by a more general action. One of the simplest possibilities of extending Einstein's gravity is to introduce an arbitrary function $f$ of the Ricci scalar $R$ into the gravitational action \cite{fR1,fR2}, which thus becomes $S=\int{\left(f(R)/2\kappa ^2+L_m\right)\sqrt{-g}d^4x}$. In this framework, a geometric solution to the dark matter problem can also be obtained \cite{fR3}. A second approach to extend the Hilbert-Einstein action is to assume the existence of a non-minimal coupling between geometry and matter. This direction of research leads to distinct classes of gravitational theories, called $f\left(R,L_m\right)$ gravity \cite{fLm1,fLm2, fLm3,fLm4}, with action given by $S=\int{f\left(R,L_m\right)\sqrt{-g}d^4x}$, and to the $f(R,T)$ gravity theory \cite{fRT1}, with action given by $S=\int{f\left(R,T\right)\sqrt{-g}d^4x}$, where $T$ is the trace of the energy-momentum tensor, respectively. Another theoretical approach, called hybrid metric-Palatini gravity, and which combines both the metric and Palatini formalisms of modified gravity theories was proposed  in \cite{hyb1,hyb2} to construct a new type
of gravitational Lagrangian. For extensive reviews and discussions of the modified gravity theories and of their implications see \cite{rev1, rev2,rev3,rev4,rev5,rev6,rev7,rev8,rev9,rev10,rev11,rev12,book}.

The properties as well as the astrophysical and cosmological implications of the $f(R,T)$ gravity theory have been investigated in detail \cite{p1,p2,p3,p4,p5,p6,p7,p8,p9,p10,p11,p12,p13,p14,p15,p16,p17,p18,p19,p20,p21,p22,p23,p24,p25,p26,p27,p28,p29,p30,p31,p32,p33}. An interesting feature of the theory is its possible interpretation as an effective description of some quantum gravity phenomena. y. As suggested in \cite{p1}, by adopting a nonperturbative approach for the quantization of the gravitational metric, proposed \cite{p2,p3,p4}, a particular type of $f(R, T )$ gravity naturally comes up due to the quantum fluctuations of the metric tensor, with the action given by $S=\int{\left(1-\alpha\right)R/2\kappa ^2+\left(L_m-\alpha T/2\right)\sqrt{-g}d^4x}$, where $\alpha $ is a constant. This interesting theoretical result may imply  the existence of a deep connection  between the quantum field theoretical description of the gravitational interaction in curved backgrounds, which
automatically involves particle creation in the gravitational field, and the corresponding effective classical description within the framework of
the $f(R, T )$ gravity theory \cite{p4}.

Since general relativity is basically a geometric theory, formulated in the Riemann metrical space, a second promising approach for obtaining generalized theories of gravity consists in looking for more general geometric structures that could describe the gravitational field.  Hence more  general geometries than the Riemannian one, which may be valid at the Solar System level only, may provide  an explanation of the behavior at large cosmological scales  of the matter in the  Universe.

The first attempt to create a more general geometry than the Riemannian one is due to Weyl \cite{Weyl}, which is a classic example of the fruitful interplay between mathematics and physics. The main goal of the study by Weyl was to obtain a geometrical unification of electromagnetism and gravitation. The fundamental concept in Riemann geometry is the metric-compatible Levi-Civita connection, which allows the comparison of lengths. Weyl did replace the metric field by the class of all conformally equivalent metrics, and he did introduce a connection that does not contain any information about the length of a vector in the parallel transport. In order to obtain information on the vector length, Weyl introduced an extra connection, the length connection, which does not contain any knowledge about the direction of a vector on parallel transport. The only role of the length connection is to  fix, or gauge, the conformal factor. The covariant divergence of the metric tensor is non-zero in Weyl's theory, and this property can be expressed mathematically in terms of a new geometric quantity, called non-metricity. In the physical applications of this geometry the length connection was identified with the electromagnetic potential. Dirac \cite{Dirac} proposed a generalization of Weyl's theory, which is based on the idea of the existence of two metrics, the physically  undetectable  metric $ds_E$, altered by the transformations in
the standards of length, while the second metric, a measurable one, is given by the conformally invariant atomic metric $ds_A$. Weyl's theory has a remarkable intrinsic mathematical beauty, associated with a rich physical structure. However, it was largely ignored by physicists, and it did not become a mainstream research topics mainly because of Einstein's very early criticism \cite{Eincr} that "...in Weyl's theory the frequency of spectral lines would
depend on the history of the atom, in complete contradiction to known experimental facts."

However, another important development in geometry, which led to a new class of generalized geometric theories of gravity,  took place  due to the work of Cartan, who, based on his geometric work \cite{Car1},  proposed an extension of general relativity \cite{Car2,Car3,Car4}, known today as the Einstein-Cartan theory \cite{Hehl1}. The torsion field, representing the new geometric element of the theory,  is usually interpreted,  from a physical point of view, as the spin density \cite{Hehl1}. The Weyl geometry can be naturally extended to include the torsion. The resulting geometry is called the Weyl-Cartan geometry, and it was widely studied from both mathematical and physical points of view \cite{WC1,WC2,WC3,WC4,WC5,WC6,WC7,WC8, WC9}. Torsion was included in the geometric framework of the Weyl-Dirac theory in \cite{Isr1, Isr2,Isr3}, leading to  an action
integral from which one can construct a general relativistic massive electrodynamics, gauge covariant in the sense of Weyl.
For a review of the of the physical applications and geometric properties of the Riemann-Cartan and Weyl-Cartan geometries see \cite{Rev}.

A third independent mathematical development that quickly did find important physical applications  took place through  the work of Weitzenb\"{o}ck \cite{Weitz}, who introduced what are presently known as the Weitzenb\"{o}ck spaces. A Weitzenb\"{o}ck manifold is characterized by the properties $\nabla _{\mu }g_{\sigma \lambda }= 0$, $T^{\mu }_{\sigma \lambda }\neq 0$, and $R^{\mu }_{\nu \sigma \lambda }=0$, where $g_{\sigma \lambda }$, $T^{\mu }_{\sigma \lambda }$ and $R^{\mu }_{\nu \sigma \lambda }$ are the metric tensor, the torsion tensor, and the curvature tensor of
the manifold, respectively. When $T^{\mu }_{\sigma \lambda }= 0$, the  Weitzenb\"{o}ck manifold is reduced to a Euclidean manifold. The torsion tensor has different values  on different regions of the Weitzenb\"{o}ck manifold. Since the Riemann curvature tensor of a  Weitzenb\"{o}ck space is zero, these  geometries have the important property of distant parallelism,  a property also known as absolute parallelism, or teleparallelism. Weitzenb\"{o}ck type space-times  were first applied in physics by Einstein, who proposed a unified teleparallel theory of electromagnetism and gravity \cite{Ein}.

In the teleparallel approach to gravity the basic idea is to replace  the metric $g_{\mu \nu}$ of the space-time, the basic physical variable describing the gravitational properties, by a set of tetrad vectors $e^i_{\mu }$. The torsion, generated by the tetrad fields, can then be used to entirely describe gravitational effects, with the  curvature replaced by the torsion. Thus we arrive to the so-called teleparallel equivalent of General Relativity (TEGR),
which was initially introduced in \cite{TE1,TE2,TE3}, and is also known presently as the $f(T)$ gravity theory. Hence, in teleparallel, or $f(T)$ type theories,   torsion exactly compensates curvature, with the important consequence that the space-time becomes flat. An important advantage of the $f(T)$ gravity theory is that the field equations are of second order, unlike in $f(R)$ gravity, which in the metric approach is a fourth order theory. For a detailed discussion of teleparallel theories see \cite{book1}.  $f(T)$ gravity theories have been widely applied to the study of astrophysical processes, and to cosmology, and in particular they are extensively used to explain the late-time accelerating expansion of the Universe, without the need of introducing dark energy \cite{TE4,TE5,TE6,TE7,TE8,TE9,TE10,TE11,TE12,TE13,TE14,TE15,TE16,TE17,TE18,TE19,TE20,TE21,TE22}.

In \cite{WCW} an extension of the teleparallel gravity models, called WCW gravity, was proposed. In this theory, the Weitzenb\"{o}ock condition of the vanishing of the sum of the curvature and torsion scalar is imposed in a background  Weyl-Cartan type space-time. A basic difference with the
standard teleparallel theories is that this the model is formulated in a four-dimensional curved space-time, and not in a flat Euclidian geometry. WCW gravity  leads to a purely geometrical description of dark energy, with the late time acceleration of the Universe fully determined by the intrinsic properties of the space-time.  An extension of the Weyl-Cartan-Weitzenb\"{o}ck (WCW) and teleparallel gravity in which the Weitzenb\"{o}ck condition of the exact cancellation of curvature and torsion in a Weyl-Cartan geometry is inserted into the gravitational action via a Lagrange multiplier was considered in \cite{WCW1}.  As a particular model the case of the Riemann-Cartan space-times with zero nonmetricity, which mimics the teleparallel theory, was considered. Several classes of exact cosmological models were also investigated.

From the above presentation it turns out that general relativity can be represented in (at least) two equivalent geometric representations: the curvature representation (in which the torsion and the nonmetricity vanish), and the teleparallel representation (in which the curvature and the nonmetricity vanish), respectively. However, a third equivalent representation is also possible, in which the basic geometric variable describing the properties of the gravitational interaction is represented by the nonmetricity $Q$ of the metric, which geometrically describes the variation of the length of a vector in the parallel transport. Such an approach, called symmetric teleparallel gravity, was initially introduced in \cite{Nester},  and it has the advantages of covariantizing the usual coordinate calculations in general relativity. It turns out that in symmetric teleparallel gravity the associated energy-momentum density is essentially the Einstein pseudotensor, which becomes a true tensor in this geometric representation. Symmetric teleparellel gravity was further developed into  the $f(Q)$ gravity theory (or coincident general relativity) in \cite{Lav}, and it is also known as nonmetric gravity. Different geometrical and physical aspects of symmetric teleparallel gravity have been investigated in the past two decades in a number of studies, with the interest for this theory increasing rapidly recently \cite{s1,s2,s3,s4,s5,s6,s7,s8,s9,s10,s11, s11a, s12,s13,s14,s15,s15a,s16,s17,s18}. For a review of teleparallel gravity see \cite{revs}.

The propagation of gravitational waves  in various extensions of symmetric teleparallel gravity was investigated in \cite{s9}, with a particular  focus on their speed and polarization. For the simple symmetric teleparallel gravity, and for theories that arise from the generalized irreducible decomposition of symmetric teleparallel gravity, as well as for $f(Q)$ gravity, the same speed and polarizations of the gravitational waves were obtained as in general relativity.  A derivation of the exact propagator for the most general infinite-derivative, even-parity and generally covariant theory in the symmetric teleparallel spacetimes was presented in \cite{s10}. In this approach the action made up of the non-metricity tensor and its contractions was decomposed into terms involving the metric and a gauge vector field. The propagation velocity of the gravitational waves around Minkowski spacetime and their potential polarizations in a general class of symmetric teleparallel gravity theories, called "newer general relativity" class, was investigated in \cite{s13}. The theory is defined in terms of the most general Lagrangian that is quadratic in the nonmetricity tensor, does not contain its derivatives and is determined by five free parameters. As a result of this investigation it was found that all gravitational waves propagate with the speed of light. The Noether Symmetry Approach was used to classify all possible quadratic, first-order derivative terms of the non-metricity tensor in the framework of Symmetric Teleparallel Geometry in \cite{s14}. The considered models were invariant under point transformations in a cosmological background. The symmetries of these models were used to reduce the dynamics of the system in order to find analytical solutions. The cosmology of the $f(Q)$ theory and its observational constraints were investigated in \cite{s15} and \cite{s15a}, and it was shown that in this theory the accelerating expansion is an intrinsic property of the  geometry of the Universe, without need of either exotic dark energy or extra fields. The dynamical system method was used to investigate the general properties of the cosmological evolution. The behaviour of the cosmological perturbations in $f(Q)$ gravity was investigated in \cite{s17}. Tensor perturbations feature a re-scaling of the corresponding Newton's constant, while vector perturbations do not contribute in the absence of vector sources. In the scalar sector two additional propagating modes were found, indicating that $f(Q)$ theories introduce, at least, two additional degrees of freedom.

An extension of symmetric teleparallel gravity was considered in \cite{s12} by introducing, in the framework of the metric-affine formalism, a new class of theories where the nonmetricity $Q$ is nonminimally coupled to the matter Lagrangian. A Lagrangian of the form $L=f_1(Q)+f_2(Q)L_m$ was considered, where $f_1$ and $f_2$ are generic functions of $Q$, and $L_m$ is the matter Lagrangian. This nonminimal coupling leads to the nonconservation of the energy-momentum tensor, and consequently the appearance of an extra force in the geodesic equation of motion. Several cosmological applications were considered for some specific functional forms of the functions $f_1(Q)$ and $f_2(Q)$, such as power-law and exponential dependencies of the nonminimal couplings. The cosmological solutions lead to  accelerating evolutions at late times.

It is the main goal of our present investigation to consider another extension of $f(Q)$ gravity, which is based on the nonminimal coupling between the nonmetricity $Q$ and the trace $T$ of the matter energy-momentum tensor. More exactly, we assume that the Lagrangian density of the gravitational field is given by a general function of both $Q$ and $T$, so that $L=f(Q,T)$. From this gravitational Lagrangian the geometric action can be constructed in the usual way. By varying the action with respect to the metric tensor we obtain the general field equations describing gravitational phenomena in the presence of geometry-matter coupling. By considering the covariant derivative of the field equations we obtain the basic result that the divergence of the  matter energy-momentum tensor does not vanish in the present approach to the gravitational interaction.  The cosmological implications of the $f(Q,T)$ theory  are investigated for three classes of specific models. The  obtained solutions describe both accelerating and decelerating evolutionary phases of the Universe, and they indicate that $f(Q,T)$ gravity can provide useful insights for the description of the early and late phases of cosmological evolution.

The  present paper is organized as follows. The geometric background, the gravitational action, the field equations and the divergence of the matter energy-momentum tensor are presented in Section~\ref{sect1}.  The cosmological formalism of $f(Q,T)$ gravity is investigated, for a homogeneous and isotropic flat geometry  in Section~\ref{sect2}. Three specific cosmological models, corresponding to different choices of the function  $f(Q,T)$, are analyzed in detail in Section~\ref{sect3}. We discuss and conclude our results in Section~\ref{sect4}.  The explicit calculations of the geometric and physical quantities necessary to obtain the field equations and the divergence of the matter energy-momentum tensor (the general expression of $Q$, the variation $\delta Q$, the variation of the gravitational action with respect to the connection, the divergence of the field equations, and the expression of $Q$ for the cosmological case) are presented in detail in Appendices~\ref{app1}-\ref{app5}.

\section{Field equations of $f(Q,T)$ theory}\label{sect1}

In the present Section we briefly review the geometrical foundations of the gravitational theories based on the assumption of the existence of a general line element in the space-time. Then we will introduce the variational principle of the $f(Q,T)$ gravitational theory, and we obtain the gravitational field equations of this geometric approach to the gravitational phenomena. The divergence of the matter energy-momentum tensor is also considered, and we show that due to the coupling between matter and geometry this tensor is not conserved.

\subsection{Geometrical preliminaries}\label{sect2}

Weyl introduced an important generalization of the Riemannian geometry, representing the mathematical basis of general relativity, by assuming that
during the parallel transport around a closed path, an arbitrary vector will not only
be subjected to a change of its direction, but it will also experience a modification of its length
\cite{Weyl}. To describe mathematically these two simultaneous changes,
Weyl proposed the introduction of a new vector field $w^{\mu }$, which, together with the metric tensor
$g_{\mu \nu}$, represent the fundamental fields of the Weyl geometry.
The Weyl geometric theory has the important characteristic  that the mathematical properties of the vector $w^{\mu }$ exactly coincide
 with those of the electromagnetic potentials. This suggests that the
 electromagnetic and gravitational forces, both long-range forces, may have a common geometric origin
\cite{Dirac}.

If in a Weyl space a vector of length $l$ is carried along an
infinitesimal path $\delta x^{\mu }$ by parallel transport, the variation in its
length $\delta l$ is given by the expression $\delta l=lw_{\mu }\delta x^{\mu }$ \cite{Dirac}. After the parallel transport of a vector
 around a small closed loop of area $\delta s^{\mu \nu}$, the
variation  of the length of the vector is given by the expression $\delta l=lW_{\mu \nu}\delta s^{\mu \nu}$, where we have denoted
\be
W_{\mu\nu}=\nabla_\nu w_{\mu}-\nabla_\mu w_{\nu},
\ee
and where the covariant derivative $\nabla_\nu$  is defined with respect to the metric $g_{\mu \nu}$.

By performing a local scaling of lengths of the form $\tilde{l}=\sigma (x)l$, the field $w_{\mu }$
changes as $\tilde{w}_{\mu }=w_{\mu }+\left(\ln \sigma \right)_{,\mu }$,
while the metric tensor coefficients are modified according to the conformal
transformations  $\tilde{g}_{\mu \nu }=\sigma ^2g_{\mu \nu}$ and $\tilde{g}^{\mu
\nu }=\sigma ^{-2}g^{\mu \nu}$, respectively \cite{Rev}. Another important property
of the Weyl geometry is the existence of the semi-metric connection
\be\label{con}
\bar{\Gamma}^{\lambda}_{~\mu\nu}=\Gamma^{\lambda}_{~\mu\nu}+g_{\mu\nu}w^{\lambda
}-\delta^{\lambda}_{\mu}w_{\nu}-\delta^{\lambda}_{\nu}w_{\mu}
,
\ee
where $\Gamma^{\lambda}_{~\mu\nu}$ denotes the usual Christoffel symbol, obtained  with the help of the
metric $g_{\mu\nu}$. In the Weyl geometry $ \bar{\Gamma}^{\lambda}_{~\mu\nu}$ is
assumed to be symmetric in its lower indices, and with its help one can construct a gauge covariant
derivative in the standard way \cite{Rev}. By using the covariant derivative one can obtain the Weyl curvature tensor, which can be written as
\be
\bar{R}_{\mu \nu \alpha \beta }=\bar{R}_{(\mu \nu )\alpha \beta }+\bar{R}_{[\mu
\nu ]\alpha \beta },
\ee
where we have defined the quantities
\bea\label{eq}
\hspace{-0.5cm}\bar{R}_{[\mu \nu ]\alpha \beta }&=&R_{\mu\nu\alpha\beta}+2\nabla_\alpha
w_{[\mu}g_{\nu]\beta}+2\nabla_\beta w_{[\nu}g_{\mu]\alpha}+\nonumber\\
\hspace{-0.5cm}&&2w_\alpha
w_{[\mu}g_{\nu]\beta}+2w_\beta
w_{[\nu}g_{\mu]\alpha}-2w^2g_{\alpha[\mu}g_{\nu]\beta},
\eea
and
\be
\bar{R}_{(\mu \nu )\alpha \beta }=\frac{1}{2}\left(\bar{R}_{\mu \nu \alpha \beta
}+\bar{R}_{\nu \mu \alpha \beta }\right)=g_{\mu \nu}W_{\alpha \beta },
\ee
respectively, with the square brackets denoting anti-symmetrization. For the first contraction of the
Weyl curvature  tensor we find
\bea
\bar{R}_{~\nu }^{\mu }&=&\bar{R}_{\;\;\;\alpha \nu }^{\alpha \mu }=R_{~\nu }^{\mu
}+2w^{\mu }w_{\nu }+3\nabla _{\nu }w^{\mu }-\nabla _{\mu }w^{\nu }+\nonumber\\
 &&g^{\mu }_{~\nu }\left(\nabla _{\alpha }w^{\alpha }-2w_{\alpha }w^{\alpha
}\right),
\eea
where by $R^\mu_{~\nu}$ we have denoted the Ricci tensor constructed from the metric. Finally, for the Weyl
scalar we obtain the expression
\be
\bar{R}=\bar{R}_{~\alpha }^{\alpha }=R+6\left(\nabla _{\mu }w^{\mu }-w_{\mu
}w^{\mu }\right).
\ee

The Weyl geometry can be generalized by taking into account the torsion of the space-time, thus obtaining the Weyl-Cartan
spaces with torsion. In a Weyl-Cartan space-time we can introduce a symmetric metric tensor
$g_{\mu \nu}$, which defines the length of a vector,  and an asymmetric connection
$\hat{\Gamma }_{~\mu \nu}^{\lambda }$, which determines the law of the parallel
transport as $dv^{\mu }=-v^{\sigma } \hat{\Gamma }_{~\sigma \nu}^{\mu }dx^{\nu
}$ \cite{Hehl1, Rev}. In the case of the Weyl-Cartan geometry the connection can be decomposed into three
irreducible parts as follows:  the Christoffel symbol $\Gamma _{~\mu \nu}^{\lambda }$, the contortion tensor $C_{~\mu
\nu}^{\lambda }$, and the disformation tensor $L_{~\mu
\nu}^{\lambda }$, respectively, so that generally one can write \cite{Hehl1}
\be\label{23}
\hat{\Gamma }_{~\mu \nu}^{\lambda }=\Gamma _{~\mu \nu}^{\lambda }+C_{~\mu
\nu}^{\lambda }+L_{~\mu \nu}^{\lambda}.
\ee

The first term in the above equation, the Levi-Civita connection of the metric $g_{\mu \nu}$, is given by its standard definition
\be
\Gamma _{~\mu \nu}^{\lambda }=\frac{1}{2}g^{\lambda \sigma} \left(\frac{\partial g_{\sigma \nu}}{\partial x^{\mu}}+\frac{\partial g_{\sigma \mu}}{\partial x^{\nu}}-\frac{\partial g_{\mu \nu}}{\partial x^{\sigma}}\right).
\ee

The contorsion tensor $C_{~\mu \nu}^{\lambda }$ in Eq. ~(\ref{23}) can be obtained from the torsion tensor $\hat{\Gamma }_{~[\mu \nu ]}^{\lambda }$, defined as
\be
\hat{\Gamma }_{~[\mu \nu ]}^{\lambda }=\frac{1}{2}\left(\hat{\Gamma }_{~\mu \nu
}^{\lambda }-\hat{\Gamma }_{~\nu \mu }^{\lambda }\right),
\ee
according to the following relation
\be
C_{~\mu \nu}^{\lambda }=\hat{\Gamma }_{~[\mu \nu ]}^{\lambda }+g^{\lambda \sigma
}g_{\mu \kappa }\hat{\Gamma }_{~[\nu \sigma ]}^{\kappa }+g^{\lambda \sigma
}g_{\nu \kappa }\hat{\Gamma }_{~[\mu \sigma ]}^{\kappa }.
\ee
As one can see immediately from the above equation, the contorsion tensor is antisymmetric with respect to its first two indices. The disformation tensor is obtained from the nonmetricity as
\be
L_{~\mu \nu}^{\lambda }=\frac{1}{2}g^{\lambda \sigma }\left(Q_{\nu \mu \sigma }+Q_{\mu
\nu \sigma}-Q_{\lambda \mu \nu }\right).
\ee

As for the non-metricity tensor $Q_{\lambda \mu \nu }$, it is defined as (minus) the
covariant derivative of the metric tensor with respect to the Weyl-Cartan connection  $\hat{\Gamma }_{~\mu
\nu}^{\lambda }$, $\nabla _{\sigma}g_{\mu \nu}=Q_{\sigma \mu \nu}$, and it can be obtained as \cite{Hehl1},
\be\label{con1}
Q_{\lambda \mu \nu}=-\frac{\partial g_{\mu \nu}}{\partial x^{\lambda }}+g_{\nu
\sigma }\hat{\Gamma }_{~\mu \lambda }^{\sigma }+g_{\sigma \mu }\hat{\Gamma
}_{~\nu \lambda }^{\sigma }.
\ee

The comparison of Eqs.~(\ref{con}) and (\ref{23}) immediately show that the Weyl
geometry is a particular case of the Weyl-Cartan geometry, in which the torsion
is zero, and the non-metricity is represented by the expression $Q_{\lambda \mu \nu }=-2g_{\mu \nu
}w_{\lambda }$. Therefore in a Weyl-Cartan geometry the connection can be written in the form
\be\label{con3}
\hat{\Gamma}^{\lambda}_{~\mu\nu}=\Gamma^{\lambda}_{~\mu\nu}+g_{\mu\nu}w^{\lambda
}-\delta^{\lambda}_{\mu}w_{\nu}-\delta^{\lambda}_{\nu}w_{\mu}
+C^{\lambda}_{~\mu\nu},
\ee
where
\be
C^{\lambda}_{~\mu \nu}=T^{\lambda}_{~\mu \nu}-g^{\lambda \beta}g_{\sigma
\mu}T^{\sigma}_{~\beta\nu}
-g^{\lambda \beta}g_{\sigma \nu}T^{\sigma}_{~\beta\mu},
\ee
is the contortion, while the Weyl-Cartan torsion
$T^{\lambda}_{~\mu \nu}$ is defined according to
\be
T^{\lambda}_{~\mu \nu}=\frac{1}{2}\left(\hat{\Gamma}^{\lambda}_{~\mu
\nu}-\hat{\Gamma}^{\lambda}_{~\nu \mu}\right).
\ee
With the use of the connection, one can define the Weyl-Cartan curvature tensor as
\be\label{curvtens}
\hat{R}^{\lambda}_{~\mu\nu\sigma}=\hat{\Gamma}^{\lambda}_{~\mu\sigma,\nu}-\hat{\Gamma}
^{\lambda}_{~\mu\nu,\sigma}+\hat{\Gamma}^{\alpha}_{~\mu\sigma}\hat{\Gamma}^{
\lambda}_{~\alpha\nu}-\hat{\Gamma}^{\alpha}_{~\mu\nu}\hat{\Gamma}^{\lambda}_{
~\alpha\sigma}.
\ee
With the use of Eq.~(\ref{con3}), one can find the curvature tensor
$\hat{R}^{\lambda}_{~\mu\nu\sigma}$ in the terms of the standard Riemann tensor, plus some new terms
containing the Weyl vector, the torsion and the contortion. By contracting the resulting
curvature tensor, one can obtain the Weyl-Cartan scalar of the geometry as follows
\begin{align}\label{eq}
\hat{R}=\hat{R}^{\mu\nu}_{~~\mu\nu}&=R+6\nabla_\nu w^\nu-4\nabla_\nu T^\nu-6w_\nu
w^\nu+8w_\nu T^\nu\nonumber\\
&+T^{\mu\alpha\nu}T_{\mu\alpha\nu}+2T^{\mu\alpha\nu}T_{\nu\alpha\mu}-4T_\nu
T^\nu.
\end{align}
where we have defined $T_\mu=T^\nu_{~\mu\nu}$, and all covariant derivatives are considered with respect to the
metric.

The symmetric teleparallel gravity is a geometric description of gravity, which is fully eq1uivalent to general relativity. This equivalence can be easily proven in the so-called coincident gauge, for which $\hat{\Gamma }_{~\mu \nu}^{\lambda }\equiv 0$. Now, by imposing the condition that the connection is symmetric, the torsion tensor identically vanishes, and the Levi-Civita connection can be expressed in terms of the disformation tensor as
\be\label{rel}
\Gamma _{~\mu \nu}^{\lambda }=-L _{~\mu \nu}^{\lambda }.
\ee

On the other hand, as it is well known ro  standard general relativity, after eliminating the boundary terms in the expression of the Ricci scalar, the gravitational action can be reformulated in a (noncovariant) form as \cite{Landau}
\be
S=\frac{1}{16\pi G}\int{g^{\mu \nu}\left(\Gamma _{\sigma \mu}^{\alpha}\Gamma ^{\sigma}_{\nu \alpha}-\Gamma _{\sigma \alpha}^{\alpha}\Gamma _{\mu \nu}^{\sigma}\right)\sqrt{-g}d^4x}.
\ee

By taking into account the relation (\ref{rel}), it turns out that in the coincident gauge the gravitational action can be reformulated in terms of the disformation tensor as
\be\label{Q}
S=-\frac{1}{16\pi G}\int{g^{\mu \nu}\left(L _{\sigma \mu}^{\alpha}L ^{\sigma}_{\nu \alpha}-L _{\sigma \alpha}^{\alpha}\Gamma _{\mu \nu}^{\sigma}\right)\sqrt{-g}d^4x}.
\ee

The action given by Eq.~(\ref{Q}), called the action of the symmetric teleparallel gravity,  is thus equivalent with the standard Hilbert-Einstein action of general relativity. However, there are some fundamental differences between the two gravitational models. In the symmetric teleparallel gravity the overall geometry of the space-time is flat, due to the vanishing of he curvature tensor (\ref{curvtens}). Hence the global geometry is of Weitzenb\"{o}ck type. Moreover, the gravitational effects are carried out not because of the rotation of the angle between two vectors in the parallel transport, but because of the variation of the length of the vector itself.

\subsection{The variational principle and the field equations of $f(Q,T)$ gravity}

In the following we will  consider an extension of the Lagrangian (\ref{Q}) of the symmetric teleparallel gravity, given by
\begin{eqnarray}\label{eq:f(Q,T)action}
S=\int \lb \frac{1}{16 \pi} f(Q, T) + \mathcal{L}_{M}\rb \sqrt{-g}\ \md ^4 x,
\end{eqnarray}
where $g\equiv \det\lp g_{\mu \nu} \rp$,  and we have defined
\begin{eqnarray}\label{eq:Q}
Q\equiv -g^{\mu \nu}\lp L^{\alpha}_{\ \ \beta\mu}L^{\beta}_{\ \ \nu\alpha} - L^{\alpha}_{\ \ \beta\alpha} L^{\beta}_{\ \ \mu \nu}  \rp,
\eea
and
\bea\label{eq:disformation}
L^{\alpha}_{\ \ \beta \gamma}\equiv -\frac{1}{2} g^{\alpha \lambda}\lp  \nabla_{\gamma}g_{\beta \lambda} + \nabla_{\beta}g_{\lambda\gamma}-\nabla_{\lambda}g_{\beta \gamma}  \rp ,
\end{eqnarray}
respectively. By $T$ we have denoted the trace of the energy-momentum tensor.
We define the trace of nonmetricity tensor as
\begin{eqnarray}
Q_{\alpha}\equiv Q_{\alpha\ \ \ \mu}^{\ \ \mu}, \ \ \ \ \tilde{Q}_{\alpha}\equiv Q^{\mu}_{\ \  \alpha \mu}.
\end{eqnarray}

We also introduce the superpotential of our model, defined  as
\begin{eqnarray}\label{eq:superpotential}
\hspace{-0.5cm}&& P^{\alpha}_{\ \ \mu\nu}\equiv \frac{1}{4}\bigg[ -Q^{\alpha}_{\ \ \mu \nu}+ 2Q^{\ \ \ \alpha}_{\lp \mu \ \ \ \nu \rp} + Q^{\alpha}g_{\mu \nu} - \tilde{Q}^{\alpha}g_{\mu\nu}\nonumber \\
\hspace{-0.5cm}&&- \delta^{\alpha}_{\ \ ( \mu} Q _{\nu )} \bigg] = -\frac{1}{2}L^{\alpha}_{\ \ \mu\nu}+ \frac{1}{4}\lp Q^{\alpha} - \tilde{Q}^{\alpha}  \rp g_{\mu \nu} - \frac{1}{4} \delta^{\alpha}_{\ \ (\mu} Q_{\nu)}. \nonumber \\
\end{eqnarray}

Then, as explicitly shown in Appendix~\ref{app1}, we obtain for $Q$ the relation
\begin{eqnarray}
&&Q=-Q_{\alpha\mu\nu}P^{\alpha\mu\nu}=- \frac{1}{4}\big( -Q^{\alpha\nu\rho}Q_{\alpha\nu\rho}+ 2 Q^{\alpha\nu\rho} Q_{\rho\alpha\nu} \nonumber \\
&&- 2Q^{\rho}\tilde{Q}_{\rho} + Q^{\rho}Q_{\rho}  \big).
\end{eqnarray}

Next, we vary the action in Eq.(\ref{eq:f(Q,T)action}) with respect to the components of the metric tensor. Hence, as a first step, we obtain,
\begin{eqnarray}
&&\delta S= \int  \frac{1}{16 \pi}\delta \lb f(Q, T) \sqrt{-g} \rb +\delta \lb \mathcal{L}_{M} \sqrt{-g} \rb \ \md ^4 x \nonumber \\
&& = \int \frac{1}{16\pi} \bigg(  -\frac{1}{2} f g_{\mu\nu} \sqrt{-g}\delta g^{\mu\nu}+f_{Q}\sqrt{-g}\delta Q   \nonumber \\
&&+ f_{T}\sqrt{-g}\delta T\bigg)-\frac{1}{2}  T_{\mu\nu}\sqrt{-g}\delta g^{\mu\nu}\ \md ^4 x,
\end{eqnarray}

The explicit form of the variation of  $\delta Q$ is presented in Appendix \ref{app2}. Moreover, as usual, we define
\begin{eqnarray}
T_{\mu\nu}\equiv -\frac{2}{\sqrt{-g}}\frac{\delta \lp \sqrt{-g}\mathcal{L}_M \rp}{\delta g^{\mu\nu}}, \ \ \ \ \Theta_{\mu\nu}\equiv g^{\alpha\beta}\frac{\delta T_{\alpha\beta}}{\delta g^{\mu\nu}},
\end{eqnarray}
which means that $\delta T= \delta (T_{\mu\nu} g^{\mu\nu})=\lp T_{\mu\nu} + \Theta_{\mu\nu} \rp \delta g^{\mu\nu}$. Then we can easily find for the variation of the action the expression
\begin{eqnarray}
&&\delta S = \nonumber \\
&&  \int \frac{1}{16\pi} \bigg\{  -\frac{1}{2} f g_{\mu\nu} \sqrt{-g}\delta g^{\mu\nu} + f_{T}\lp T_{\mu\nu} + \Theta_{\mu\nu} \rp \sqrt{-g} \delta g^{\mu\nu} \nonumber \\
&&-f_{Q}\sqrt{-g} \lp P_{\mu\alpha\beta}Q_{\nu}^{\ \ \alpha\beta} -2 Q^{\alpha\beta}_{\ \ \ \ \mu}P_{\alpha\beta\nu} \rp \delta g^{\mu\nu}\nonumber \\
&&+2 f_{Q}\sqrt{-g} P_{\alpha\mu\nu} \nabla^{\alpha}\delta g^{\mu\nu} \bigg \} -\frac{1}{2}  T_{\mu\nu}\sqrt{-g}\delta g^{\mu\nu}\ \md ^4 x .
\end{eqnarray}

As for the term $2 f_{Q}\sqrt{-g} P_{\alpha\mu\nu} \nabla^{\alpha}\delta g^{\mu\nu}$, after integration and with the use of the boundary conditions it turns out that it takes the form $-2 \nabla^{\alpha}\lp f_{Q}\sqrt{-g} P_{\alpha\mu\nu} \rp \delta g^{\mu\nu}$. Finally, after equating the variation of the gravitational action to zero, we obtain the field equations of the $f(Q,T)$ gravity theory as,
\begin{eqnarray}\label{eq:EOMs}
&&-\frac{2}{\sqrt{-g}}\nabla_{\alpha}\lp f_{Q}\sqrt{-g} P^{\alpha}_{\ \ \mu\nu}   \rp - \frac{1}{2}f g_{\mu\nu} + f_{T}\lp T_{\mu\nu} + \Theta_{\mu\nu} \rp \nonumber \\
&&  -f_{Q}\lp P_{\mu\alpha\beta}Q_{\nu}^{\ \ \alpha\beta} -2 Q^{\alpha\beta}_{\ \ \ \ \mu}P_{\alpha\beta\nu} \rp = 8\pi T_{\mu\nu}.
\end{eqnarray}

Ref.~\cite{s12} also has similar terms like Eq.~(\ref{eq:EOMs}), even that the considered basic physical model and action are somehow different from the present approach.

By using the Lagrangian Multiplier Method with two constrains $T^{\alpha}_{\ \ \beta \gamma}=0$ and $R^{\alpha}_{\ \ \beta\mu\nu}=0$, we can find the variation with respect to the connection. The explicit calculations are presented in Appendix \ref{app3}.  Moreover, we define the hypermomentum tensor density as
\begin{eqnarray}\label{eq:hypermomentum}
H_{\lambda}^{\ \ \mu\nu}\equiv \frac{\sqrt{-g}}{16\pi}f_{T}\frac{\delta T}{\delta \hat{\Gamma}^{\lambda}_{\ \ \mu\nu}} +\frac{\delta \sqrt{-g}\mathcal{L}_{M}}{\delta \hat{\Gamma}^{\lambda}_{\ \ \mu\nu}}.
\end{eqnarray}

By taking into account the anti-symmetry property of $\mu$ and $\nu$ in the Lagrangian multiplier coefficients $\lambda_{\alpha}^{\ \ \mu\nu}$ and $\xi_{\alpha}^{\ \ \beta\mu\nu} $, we can eliminate them by introducing $\nabla_{\mu}\nabla_{\nu}$ into the original part of action variation. Hence, after taking the variation of the gravitational action with respect to the connection we obtain the  field equations
\begin{eqnarray}\label{eq:feqconnection}
\nabla_{\mu}\nabla_{\nu}\bigg( \sqrt{-g}f_{Q} P^{\mu\nu}_{\ \ \ \ \alpha} +4\pi H_{\alpha}^{\ \ \mu \nu} \bigg)=0.
\end{eqnarray}

\subsection{The energy-momentum tensor balance  equation}

For a (1,1)-form tensor $v^{\mu}_{\ \nu}$ we define its covariant derivative as
\begin{eqnarray}
 &&\gra_{\mu}v^{\mu}_{\ \ \nu}=\p_{\mu}v^{\mu}_{\ \  \nu}+\hat{\Gamma}^{\mu}_{\ \ \mu\rho}v^{\rho}_{\ \ \nu}-\hat{\Gamma}^{\rho}_{\ \ \mu\nu}v^{\mu}_{\ \ \rho}\nonumber \\
 &&=\p_{\mu}v^{\mu}_{\ \ \nu}+\left\{^{\mu}_{\ \ \mu\rho}\right\}v^{\rho}_{\ \ \nu}+L^{\mu}_{\ \ \mu\rho}v^{\rho}_{\ \ \nu}-\left\{^{\rho}_{\ \ \mu\nu}\right\}v^{\mu}_{\ \ \rho}-L^{\rho}_{\ \ \mu\nu}v^{\mu}_{\ \ \rho}\nonumber\\
 &&=\D_{\mu}v^{\mu}_{\ \ \nu}+L^{\mu}_{\ \ \mu\rho}v^{\rho}_{\ \ \nu}-L^{\rho}_{\ \ \mu\nu}v^{\mu}_{\ \ \rho}\nonumber \\
 &&=\D_{\mu}v^{\mu}_{\ \ \nu}-\frac{1}{2}Q_{\rho}v^{\rho}_{\ \ \nu}-L^{\rho}_{\ \ \mu\nu}v^{\mu}_{\ \ \rho}.
\end{eqnarray}
Here we have $\hat{\Gamma}^{\alpha}_{\ \ \mu\nu}=\Gamma^{\alpha}_{\ \ \mu\nu}+L^{\alpha}_{\ \ \mu\nu}$, while by $\Gamma ^{\alpha}_{\ \ \mu\nu}$ we have denoted the Levi-Civita connection associated to the metric. $\D_{\mu} $ denotes the covariant derivative with respect to the Levi-Civita connection. From Eq.~(\ref{eq:disformation}) one can easily check that $L^{\mu}_{\ \ \mu\rho}=-1/2 \ Q_{\rho}$ . The field equations in the (1,1)-form are given by
\begin{eqnarray}\label{eq:metricdivfeq}
&&f_{T}\big(T^{\mu}_{\ \ \nu}+\Theta^{\mu}_{\ \ \nu}\big)-8\pi T^{\mu}_{\ \ \nu}\nonumber \\
&&=\frac{f}{2}\delta^{\mu}_{\ \ \nu}+f_{Q}Q_{\nu}^{\ \ \alpha\beta}P^{\mu}_{\ \ \alpha\beta}+\frac{2}{\g}\nabla_{\alpha}\big(f_{Q}\g P^{\alpha\mu}_{\ \ \ \ \nu}\big). \nonumber \\
\end{eqnarray}

 The metric divergence of the field equations (\ref{eq:metricdivfeq}) is explicitly calculated in Appendix~\ref{app4}, and it is given by
\begin{eqnarray}\label{eq:appmetricdiv}
\hspace{-0.5cm}\D_{\mu}\Big[ f_{T}\big(T^{\mu}_{\ \ \nu}+\Theta^{\mu}_{\ \ \nu}\big)-8\pi T^{\mu}_{\ \ \nu}\Big]+\frac{8\pi}{\g}\nabla_{\alpha}\nabla_{\mu}H_{\nu}^{\ \ \alpha\mu}\nonumber \\
\hspace{-0.5cm}= \frac{1}{2}f_{T}\p_{\nu}T + \frac{1}{\g} Q_{\mu}\gra_{\alpha}\Big(f_{Q}\g P^{\alpha\mu}_{\ \ \ \ \nu}\Big).
\end{eqnarray}

To simplify the above equation, we first solve Eq.~(\ref{eq:feqconnection}) by introducing the tensor $A^{\nu}_{\ \ \alpha}$, so that
\begin{eqnarray}\label{eq:gauge}
\nabla_{\mu}\bigg( \sqrt{-g}f_{Q} P^{\mu\nu}_{\ \ \ \ \alpha} +4\pi H_{\alpha}^{\ \ \mu \nu} \bigg)=\g A^{\nu}_{\ \ \alpha},
\end{eqnarray}
where we have the additional constrain that
\begin{eqnarray}\label{eq:gaugeconserve}
\nabla_{\nu}\Big(\g A^{\nu}_{\ \ \alpha}\Big)=\frac{\g}{2}Q_{\nu}A^{\nu}_{\ \ \alpha}+\g\nabla_{\nu}A^{\nu}_{\ \ \alpha}=0.
\end{eqnarray}

 We can always add an anti-symmetrical tensor $\nabla_{\mu}M_{\alpha}^{\ \ [\mu\nu]}$ to $\g A^{\nu}_{\ \ \alpha}$ without adding extra terms to Eq.~(\ref{eq:gaugeconserve}). Now we simply combine Eq.~(\ref{eq:appmetricdiv}) and Eq.~(\ref{eq:gauge}), and we find another form of the energy-momentum balance equation, given by
\begin{eqnarray}
\D_{\mu}\Big[ f_{T}\big(T^{\mu}_{\ \ \nu}+\Theta^{\mu}_{\ \ \nu}\big)-8\pi T^{\mu}_{\ \ \nu}\Big]+\frac{16\pi}{\g}\nabla_{\alpha}\nabla_{\mu}H_{\nu}^{\ \ \alpha\mu}\nonumber\\
-8\pi\nabla_{\mu}\bigg(  \frac{1}{\g}\nabla_{\alpha}H_{\nu}^{\ \ \alpha\mu} \bigg)+2\nabla_{\mu}A^{\mu}_{\ \ \nu}= \frac{1}{2}f_{T}\p_{\nu}T,
\end{eqnarray}
or, equivalently,
\begin{eqnarray}
\hspace{-0.5cm}&&\mathcal{D}_{\mu }T_{\ \ \nu }^{\mu } =\frac{1}{f_{T}-8\pi }\Bigg[-\mathcal{D}%
_{\mu }\left( f_{T}\Theta _{\ \ \nu }^{\mu }\right) -\frac{16\pi }{\sqrt{-g}}%
\nabla _{\alpha }\nabla _{\mu }H_{\nu }^{\ \ \alpha \mu } \nonumber\\
\hspace{-0.5cm}&&+8\pi \nabla _{\mu }\bigg(\frac{1}{\sqrt{-g}}\nabla _{\alpha }H_{\nu }^{\ \
\alpha \mu }\bigg)-2\nabla _{\mu }A_{\ \ \nu }^{\mu }+\frac{1}{2}%
f_{T}\partial _{\nu }T\Bigg] =B_{\nu }. \nonumber \\
\end{eqnarray}

Hence in the $f(Q,T)$ gravity theory the matter energy-momentum tensor is
not conserved, $\mathcal{D}_{\mu }T_{\ \ \nu }^{\mu }=B_{\nu }\neq 0$, with
the  nonconservation vector a function of $Q$, $T$, and of the
thermodynamics quantities of the system. For a perfect fluid, described by
its energy density $\rho $ and its pressure $p$, respectively, the
energy-momentum tensor takes the form $T_{\ \ \nu }^{\mu }=\left( \rho
+p\right) u_{\nu }u^{\mu }+p\delta _{\nu }^{\mu }$, $u^{\mu }$ is the
four-velocity of the fluid, normalized as $u^{\mu }u_{\mu }=-1$. Then, as
shown in \cite{s12}, from the divergence of the energy-momentum tensor we
obtain the energy balance and the momentum conservation equations as
\begin{equation}
\dot{\rho}+3H\left( \rho +p\right) =B_{\mu }u^{\mu },  \label{nc1}
\end{equation}%
and
\begin{equation}
\frac{d^{2}x^{\mu }}{ds^{2}}+\Gamma _{\alpha \beta }^{\mu }u^{\alpha
}u^{\beta }=\frac{h^{\mu \nu }}{\rho +p}\left( B_{\nu }-\mathcal{D}_{\nu
}p\right) ,  \label{nc2}
\end{equation}%
respectively,   where we have denoted by an overdot the quantity $\dot{f}$ $%
=u_{\mu }\mathcal{D}^{\mu }f$, while we have defined $H=\left( 1/3\right)
\mathcal{D}^{\mu }u_{\mu }$. $h^{\mu \nu }$ is the projection operator,
given by $h^{\mu \nu }=g^{\mu \nu }+u^{\mu }u^{\nu }$.  Eq.~(\ref{nc1})
describes the energy balance in $f(Q,T)$ gravity. From a physical point of
view it gives the amount of energy that enters or goes out in a specified
volume of a physical system. The source term $B_{\mu }u^{\mu }$
corresponds to the energy creation/annihilation. The total energy of the
gravitating system is conserved only if the condition $B_{\mu }u^{\mu }=0$
is satisfied in all points of the spacetime. If $B_{\mu }u^{\mu }\neq 0$,
then energy transfer processes or particle production takes place in the
given system.

Eq. (\ref{nc2}) represents the equation of motion of massive
particles in $f(Q,T)$ gravity. As it can be seen immediately from the
equation of motion, the dynamical evolution of the massive particles  is not
geodesic, and an extra-force with components $F^{\mu }=h^{\mu \nu }\left(
B_{\nu }-\mathcal{D}_{\nu }p\right) /\left( \rho +p\right) $ does appear,
due to the coupling between $Q$ and $T$. Hence  in $f(Q,T)$ gravity a
supplementary force is exerted on any particle, besides the usual
gravitational force. $F^{\mu }$ is orthogonal to the matter four-velocity $%
u_{\mu }$, since from the properties of the projection operator it follows
that we always have $F^{\mu }u_{\mu }=0$, which is the standard requirement
for  a physical force, for which only the components that are orthogonal to
the four-velocity of the particle can contribute to its equation of motion.

\section{Cosmological evolution of the Friedmann-Lemaitre-Robertson-Walker Universe in $f(Q,T)$ gravity}\label{sect2}

We are going now to consider the cosmological applications of the $f(Q,T)$ theory, by assuming that the Universe is described by the isotropic, homogeneous and spatially flat Friedmann-Lemaitre-Robertson-Walker  (FLRW) metric, given by
\begin{eqnarray}
\md s^2 =-N^2(t)\md t^2 + a^2(t)\delta_{i j}\md x^i \md x^j,
\end{eqnarray}
where $a(t)$ is the scale factor, and the lapse function $N(t)=1$ is for the standard case. The expansion and dilation rates are defined as follows
\begin{eqnarray}
H\equiv \frac{\dot{a}}{a}, \ \ \ \ \tilde{T}\equiv  \frac{\dot{N}}{N}
\end{eqnarray}
In cosmology $H(t)$ is called the Hubble function. By adopting the coincident gauge, in  the covariant derivatives reduce to ordinary derivatives, after straightforward calculations presented in the Appendix~\ref{app5}, we find
\begin{eqnarray}
Q=6 \frac{H^2}{N^2}.
\end{eqnarray}

\subsection{The generalized Friedmann equations}

To derive the two generalized Friedmann equations describing the cosmological evolution, we assume that the matter content of the Universe consists of perfect fluid, whose energy-momentum tensor is given by $T^{\mu}_{\ \ \nu}=\mathrm{diag}(-\rho, p, p, p)$. Then for the tensor  $\Theta^{\mu}_{\ \ \nu}$ we obtain the expression
\be
\Theta^{\mu}_{\ \ \nu}=\delta^{\mu}_{\ \ \nu}p-2T^{\mu}_{\ \ \nu}=\mathrm{diag}(2\rho+p, -p, -p, -p).
\ee

To simplify the mathematical formalism we introduce the notations
\be
F\equiv f_{Q},
\ee
 and
 \be
 8\pi \tilde{G}\equiv f_{T},
 \ee
 respectively. By using the FLRW metric, from the field equations we can easily find
\begin{eqnarray}
\hspace{-1.2cm}&&\frac{f}{2}-6F\frac{H^2}{N^2}=8\pi\rho+8\pi \tilde{G}(\rho+ p),\label{eq:friedone}\\
\hspace{-1.2cm}&&\frac{f}{2}-\frac{2}{N^2}\Big[\lp\dot{F}-F\tilde{T}\rp H+F\lp\dot{H}+3H^2\rp \Big]=-8\pi p.  \label{eq:friedtwo}
\end{eqnarray}
By solving Eq.~(\ref{eq:friedone}) and Eq.~(\ref{eq:friedtwo}) we obtain
\begin{eqnarray}
&&8\pi p=-M+S, \ \ \ \
8\pi \rho= M- \frac{\tilde{G}}{1+\tilde{G}}S,\label{eq:friedall}
\end{eqnarray}
where we have denoted
\begin{eqnarray}
\hspace{-0.9cm}&&M\equiv \frac{f}{2}-6F\frac{H^2}{N^2}, \ \ \ \
S\equiv\frac{2\dot{F}H}{N^2} + \frac{2F}{N^2}\lp \dot{H}-H\tilde{T} \rp .
\end{eqnarray}

By explicitly including  $\dot{\rho}$ and $\dot{p}$ in the expression of $\dot{f}=F\dot{Q}+8\pi G\dot{T}$, we obtain the generalized energy balance equation in $f(Q,T)$ gravity as
\begin{eqnarray}\label{50}
\dot{\rho}+3H\lp \rho + p \rp=\frac{\tilde{G}}{16\pi \lp 1+\tilde{G}\rp \lp 1+2\tilde{G} \rp }  \times \nonumber \\
\lb \dot{S}-\frac{\lp 3\tilde{G}+2 \rp \dot{\tilde{G}}}{\lp 1+\tilde{G}\rp \tilde{G}}S+6HS \rb.
\end{eqnarray}

We can easily see from the above equation that when $f$ has no $T$ dependence, which means $G=0$, the continuity equation is always valid.

Next, we consider the case when $N=1$, which is the case of the standard FRW geometry. Thus we have $Q=6H^2$, $M=f/2-6FH^2$, $S=2\lp \dot{F}H+F\dot{H}\rp$, and the generalized Friedmann equations reduce to
\be\label{51}
8\pi \rho =\frac{f}{2}-6FH^2 - \frac{2\tilde{G}}{1+\tilde{G}}\lp \dot{F}H +F \dot{H} \rp,
\ee
\be\label{52}
\hspace{-1.1cm}8\pi p= -\frac{f}{2}+6FH^2+ 2\left(\dot{F}H +F \dot{H}\right).
\ee
Combining the above two equations, we obtain the evolution equation for the Hubble function $H$ as
\begin{eqnarray}\label{53}
\dot{H} + \frac{\dot{F}}{F}H =\frac{4\pi}{F} \lp 1+\tilde{G} \rp \lp \rho + p \rp.
\end{eqnarray}

We can bring the cosmological evolution equations to a form similar to the standard general relativity Friedmann's equations, by  defining an effective energy density $\rho _{eff}$ and an effective pressure $p_{eff}$ so that,
\be\label{54}
3H^2=8\pi \rho _{eff}=\frac{f}{4F}-\frac{4\pi}{F}\lb \lp 1+\tilde{G}\rp \rho + \tilde{G} p \rb,
\ee
\bea\label{55}
2\dot{H}+3H^2&=&-8\pi p_{eff}=\frac{f}{4F}-\frac{2\dot{F}H}{F}+\nonumber\\
&&\frac{4\pi}{F}\lb \lp 1+\tilde{G} \rp \rho +\lp 2+\tilde{G} \rp p \rb .
\eea
Then it follows that the effective thermodynamic quantities satisfy the conservation equation
\be
\dot{\rho}_{eff}+3H\left(\rho _{eff}+p_{eff}\right)=0.
\ee

An important cosmological quantity is the deceleration parameter $q$, which is an indicator of the accelerating/decelerating nature of the evolution of the Universe. The deceleration parameter is defined as
\begin{equation}\label{deccparam}
q=\frac{d}{dt}\frac{1}{H}-1=-\frac{\dot{H}}{H^2}-1=\frac{1}{2}\lp 1+3w\rp ,
\end{equation}
where $w=p_{eff}/\rho_{eff}$ is the parameter of the equation of state of the dark energy. Negative values of the deceleration parameter indicate an accelerating evolution, while
positive values indicate decelerating expansion.

Explicitly, the deceleration parameter  can be expressed as,
\begin{eqnarray}\label{58}
q=-1+ \frac{3\left(4\dot{F}H-f+16\pi p\right)}{f-16\pi \lb\lp1+\tilde{G} \rp \rho +\tilde{G} p \rb}.
\end{eqnarray}

To obtain cosmological
results that can allow a direct comparison of the model predictions  with the astronomical observations,
we introduce, instead of the time variable $t$,  as independent variable the
redshift $z$, defined according to
\begin{equation}
1+z=\frac{1}{a},
\end{equation}%
where we have normalized the scale factor so that its present day value is
one, $a(0)=1$. Therefore for the time operator we obtain
\begin{equation}  \label{61a}
\frac{d}{dt}=\frac{dz}{dt}\frac{d}{dz}=-(1+z)H(z)\frac{d}{dz}.
\end{equation}%
The deceleration parameter $q$ can be obtained as a function of the cosmological redshift $z$ as
\begin{equation}
q(z)=(1+z)\frac{1}{H(z)}\frac{dH(z)}{dz}-1.
\end{equation}

We will also compare the behavior of the cosmological parameters in the $f(Q,T)$ gravity with the standard $\Lambda $CDM model. We assume that the late Universe is filled with dust matter only, having negligible pressure. Then form the standard general relativistic energy conservation equation $\dot{\rho}+3H\rho=0$ we find for the variation of matter energy density the expression $\rho\sim 1/a^3\sim (1+z)^3$. The evolution of the Hubble function is given by \cite{1f}
\be
H=H_0\sqrt{\left(\Omega _{DM}+\Omega _b\right)a^{-3}+\Omega _{\Lambda}},
\ee
where $\Omega _{DM}$, $\Omega _b$ and $\Omega _{\Lambda}$ are the density parameters of the cold dark matter, baryonic matter, and dark energy (interpreted as a cosmological constant), respectively. The density parameters satisfy the important constraint $\Omega _{DM}+\Omega _b+\Omega _{\Lambda}=1$. The Hubble function $H(z)=H_0h(z)$ can be written as a function of the redshift in a dimensionless form as
\be\label{63}
h(z)=\sqrt{\left(\Omega _{DM}+\Omega _b\right)\left(1+z\right)^{3}+\Omega _{\Lambda}}.
\ee

The redshift dependence of the deceleration parameter is obtained as
\be
q(z)=\frac{3 (1+z)^3 \left(\Omega _{DM}+\Omega _b\right)}{2 \left[\Omega _{\Lambda}+(1+z)^3
   \left(\Omega _{DM}+\Omega _b\right)\right]}-1.
\ee

In the following for the density parameters we adopt the numerical values $\Omega _{DM}=0.2589$, $\Omega _{b}=0.0486$, and $\Omega _{\Lambda}=0.6911$ \cite{1f}, obtained from the Planck data, giving for the total matter density parameter $\Omega _m=\Omega _{DM}+ \Omega _b$ the numerical value $\Omega _m=0.3089$. From these numerical values of the cosmological parameters it follows that the present day value of the deceleration parameter as $q(0)=-0.5381$. As for the variation of the dimensionless matter density with respect to the redshift, we obtain the expression $r(z)=\Omega _m(1+z)^3=0.3089(1+z)^3$.

\subsection{The de Sitter solution}

Before considering specific cosmological models of $f(Q,T)$ gravity, we would to find the vacuum solution for our field equations, and check if the theory admits a de Sitter type solution, which corresponds to the constrains $\rho=p=0$ and $H=H_0=$constant, respectively. For a vacuum Universe Eq.~(\ref{eq:friedall}) suggests $M=S=0$, where $S=0$ gives a constant $F=F_0$, which implies $f=F_0 Q+\Lambda$, with $\Lambda$  also an arbitrary constant of integration. The condition $M=0$ reduces to
\begin{eqnarray}
M=\frac{f}{2}-6F H^2=\frac{\Lambda}{2} - 3F_0 H_0^2 =0,
\end{eqnarray}
which simply gives $H_0 =\sqrt{\Lambda/6F_0}$. This result is similar to the one in Ref.~ \cite{s12}, which also gives a result equivalent to the general relativistic case when $F_0=1$. Hence the $f(Q,T)$ theory admits the de Sitter type evolution in the limiting case of a vacuum Universe.  As one can easily calculate, for the de Sitter solution we have $q=-1$ and $w=-1$, respectively.

\section{Specific cosmological models}\label{sect3}

In the present Section we will investigate some specific cosmological models in the $f(Q,T)$ gravity theory, corresponding to different choices of the functional form of $f(Q,T)$. For the sake of generality we will assume that the cosmological matter satisfies an equation of state of the form $p=(\gamma -1)\rho$, where $\gamma $ is a constant, and $1\leq \gamma \leq 2$. Such a linear barotropic equation of state can describe the baryonic matter behavior in both the high density limit (corresponding to the early Universe), and in the low density limit, appropriate for the description of the present day Universe.

With the use of the barotropic equation of state, from Eqs.~(\ref{51}) and (\ref{53}) we obtain for the matter density the general expression
\be\label{66}
\rho =\frac{f-12FH^2}{16\pi \left(1+\gamma \tilde{G}\right)}.
\ee

\subsection{$f(Q,T)=\alpha Q+\beta T$}

As a first example of the cosmological evolution in $f(Q,T)$ gravity we will consider the case in which the function $f(Q,T)$ has the simple form  $f(Q,T)=\alpha Q+\beta T$, where $\alpha $ and $\beta $ are constants. Then we immediately obtain $F=F_Q=\alpha$, and $8\pi \tilde{G}=f_T=\beta$. Hence Eq.~(\ref{53}) becomes
\be\label{65}
\dot{H}=\frac{4\pi \gamma}{\alpha }\left(1+\frac{\beta }{8\pi }\right)\rho.
\ee

Eqs.~(\ref{54}) and (\ref{55}) take the form
\be
H^2=\frac{\rho  \left[\beta  (\gamma -4)-16 \pi \right]}{6 \alpha },
\ee
and
\be
H^2= \frac{\rho  \left[\beta  (9 \gamma -4)+16 \pi  (4 \gamma -1)\right]}{6 \alpha
   },
\ee
respectively, which leads to the consistency condition
\be
\frac{4 (\beta +8 \pi ) \gamma  \rho }{3 \alpha }=0.
\ee
For $\rho \neq 0$, the above condition implies $1+\beta/8\pi=0$, which in turn leads, with the use of Eq.~(\ref{65}), to the equation $\dot{H}=-$, or $H=H_0={\rm constant}$, and $a(t)=e^{H_0t}$. The cosmological evolution is of de Sitter type, in the presence of a nonvanishing matter energy density. The evolution of $\rho$ can be obtained from the conservation equation (\ref{50}), which taking into account that in the present model $S=0$, becomes
\be
\dot{\rho}+3\gamma H_0\rho=0,
\ee
giving $\rho (t)=\rho _0e^{-3\gamma H_0t}$, where $\rho _0$ is an arbitrary constant of integration. Hence, the exponential expansion of the Universe is associated, in this model of the $f(Q,T)$ gravity theory, with an exponential decrease of the matter content.

\subsection{$f(Q,T)=\alpha Q^{n+1}+\beta T$}

As a second example of a cosmological model in the $f(Q,T)$ gravity we consider the case for which the function $f(Q,T)$ is given by  $f(Q,T)=\alpha Q^{n+1}+\beta T $, where $\alpha$, $n$ and $\beta $ are constants. Then we easily obtain
\be
F=(n+1)\alpha Q^n=6^n(n+1)\alpha H^{2n}, 8\pi \tilde{G}=\beta.
\ee

Then from Eq.~(\ref{66}) we obtain the expression of the matter density as
\be
\rho =\frac{  6^{n+1} (2 n+1)\alpha H^{2 (n+1)}}{ \beta  (\gamma
   -4)-16 \pi}.
\ee
By using this expression of the density it follows that the evolution equation for $H$, Eq.~(\ref{53}), takes the simple form
\be
\dot{H}+\frac{3 (\beta +8 \pi ) \gamma  H^2}{(n+1) (16 \pi -\beta  (\gamma
   -4))}=0,
\ee
and it has the general solution
\begin{equation}
H(t)=\frac{H_{0}(n+1)\left[ 16\pi -\beta (\gamma -4)\right] }{3(\beta +8\pi
)\gamma H_{0}\left( t-t_{0}\right) -(n+1)\left[ \beta \gamma -4(\beta +4\pi )%
\right] }.
\end{equation}
where we have used the initial condition $H\left(t_0\right)=H_0$. The evolution of the scale factor is given by
\begin{equation}
a(t)=a_{0}\left[ 3(\beta +8\pi )\gamma H_{0}(t-t_{0})+a_{1}\right] ^{\frac{%
(n+1)(16\pi -\beta (\gamma -4))}{3(\beta +8\pi )\gamma }},
\end{equation}%
where $a_0$ is an arbitrary constant of integration, and we have denoted $a_{1}=(n+1)\left[ 4(\beta +4\pi )-\beta \gamma \right]
$. The deceleration parameter is constant, and it is given by
\be
q=\frac{3 (\beta +8 \pi ) \gamma }{(n+1) \left[16 \pi -\beta  (\gamma -4)\right]}-1.
\ee
If the model parameters satisfy the constraint $3 (\beta +8 \pi ) \gamma /(n+1) \left[16 \pi -\beta  (\gamma -4)\right]<1$, the deceleration parameter takes negative values, and the expansion of the Universe is accelerating.

\subsection{$f(Q,T)=-\alpha Q-\beta T^2$}

Finally, as a simple example of a cosmological model in $f(Q,T)$ gravity, we will consider the case when the function $f(Q,T)$ has the form $f(Q,T)=-\alpha Q-\beta T^2$, where $\alpha >0$ and $\beta >0$ are constants. Moreover, for the sake of simplicity, we will fix the equation of state of the cosmological matter from the beginning as dust, that is, we choose $\gamma =1$, giving $p=0$. Then we immediately obtain
\be
F=-\alpha, 8\pi \tilde{G}=-2\beta T=2\beta \rho (t).
\ee
Eq.~(\ref{66}) gives for the matter density the simple algebraic equation
\be
\rho (t)=\frac{-\beta  \rho ^2(t)+6 \alpha  H^2(t)}{16 \pi  \left[1+2 \beta  \rho (t)\right]},
\ee
which has the physical solution
\begin{equation}
\rho (t)=\frac{8\pi \left[ \sqrt{1+3(1+32\pi )\alpha \beta H^{2}(t)/32\pi ^2}-1\right]
}{\beta \left( 1+32\pi \right) }.
\end{equation}
If the condition $3(1+32\pi )\alpha \beta H^{2}(t)/32\pi ^2<<1$, is satisfied, by power expanding the square root in the above equation gives $\rho (t)\propto H^2(t)$. Thus in this limit we recover the standard general relativistic result.

The evolution equation for the Hubble function, Eq~(\ref{66}), takes for this model the form
\begin{eqnarray}\label{81}
&&\dot{H}(t) =-\frac{32\pi ^{2}}{\alpha \beta \left( 1+32\pi \right) } \times \nonumber\\
&&\Bigg\{ 1+\frac{2\left[ \sqrt{1+3(1+32\pi )\alpha \beta H^{2}(t)/32\pi ^{2}}%
-1\right] }{1+32\pi }\Bigg\} \times  \nonumber\\
&&\left[ \sqrt{1+3(1+32\pi )\alpha \beta H^{2}(t)/32\pi ^{2}}-1\right] .
\end{eqnarray}

We rescale now the Hubble function according to
\be
H(t)=H_0h(t)=\sqrt{\frac{32\pi^2}{3\lp 1+32\pi\rp\alpha \beta }}\;h(t),
\ee
\\
and we introduce the model parameter $\lambda $, defined as
\be
\lambda =\sqrt{\frac{96\pi ^2}{(1+32\pi)\alpha \beta}}.
\ee
Then, in the new variables, Eq.~(\ref{81}) becomes
\begin{equation}\label{84}
\frac{dh(t)}{dt}=-\lambda \left\{ 1+\frac{2\left[ \sqrt{1+h^{2}(t)}-1\right]
}{1+32\pi }\right\} \left[ \sqrt{1+h^{2}(t)}-1\right] .
\end{equation}

In terms of the redshift $z$ Eq. (\ref{84}) takes the form
\bea
\left( 1+z\right) h(z)\frac{dh(z)}{dz}&=&\lambda \left\{ 1+\frac{2\left[ \sqrt{%
1+h^{2}(z)}-1\right] }{1+32\pi }\right\}\times \nonumber\\
&& \left[ \sqrt{1+h^{2}(z)}-1\right] .
\eea

By introducing the new variable $u(z)=h^{2}(z)$, the above equation can be
written as
\begin{equation}\label{86}
(1+z)\frac{du}{dz}=2\lambda \left\{ 1+\frac{2\left[ \sqrt{1+u(z)}-1\right] }{%
1+32\pi }\right\} \left[ \sqrt{1+u(z)}-1\right] .
\end{equation}

In the limit $h^2(t)<<1$, Eq.~(\ref{84}) can be approximated as
\be
\frac{dh(t)}{dt}=-\frac{\lambda }{2}h^2(t),
\ee
with the general solution given by
\be
h(t)=\frac{2h_0}{2+h_0\lambda \lp t-t_0\rp},
\ee
where $h_0=h\left(t_0\right)$. From $H(t)=H_0h(t)=\dot{a}/a$ we obtain the scale factor as $a(t)=\left[2+h_0\lambda\left(t-t_0\right)\right]^{2H_0/\lambda}$. The deceleration parameter is given by $q=\lambda /2H_0-1=\sqrt{3}/2-1<0$, while the matter energy density varies as $\rho (t)=12H_0^2h_0^2/\left[2+h_0\lambda \lp t-t_0\rp\right]^2$.

\begin{figure}
 \centering
 \includegraphics[scale=0.70]{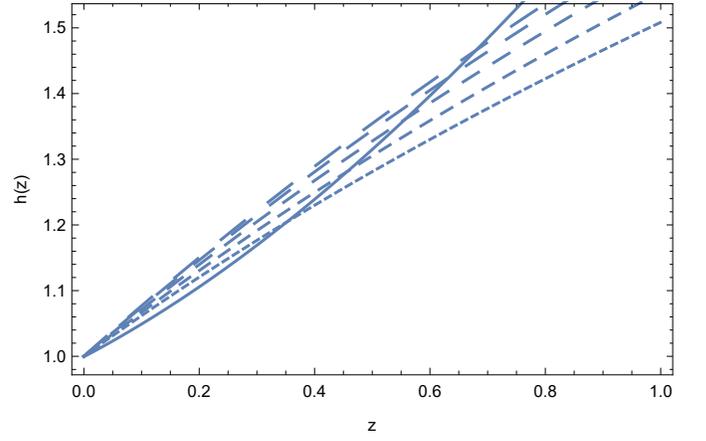}
 \caption{Variation of the Hubble function as a function of the redshift $z$ in the $f(Q,T)$ gravity theory for $f(Q,T)=-\alpha Q-\beta T^2$ and for a dust Universe,  for different values of the parameter
 $\lambda $: $\lambda =1.52$ (dotted curve), $\lambda   =1.64$ (short dashed curve),
 $\lambda =1.75$ (dashed curve), $\lambda =1.83$ (long dashed curve), and $\lambda =1.88$ (ultra long dashed curve), respectively. The variation of the Hubble function in the standard $\Lambda$CDM model is also represented as the solid curve.}
 \label{fig1}
\end{figure}

The variation of the Hubble function as a function of the redshift, obtained by numerically integrating Eq.~(\ref{86}),  is presented in Fig.~\ref{fig1}. The evolution equation for the Hubble function was integrated with the initial condition $u(0)=1$, and we have considered the redshift range $z\in[0,1]$.

As one can see from Fig.~\ref{fig1}, the Hubble function is a monotonically increasing function of the redshift (monotonically decreasing function of time) for all considered values of the model parameter $\lambda$. For small values of $z$ the cosmological evolution is practically independent on the numerical values of $\lambda$, but at higher redshifts there is a significant effect of the parameter value on the cosmic expansion. For the sake of comparison we have also presented the variation of the Hubble function in the standard $\Lambda $CDM model, as given by Eq.~(\ref{63}). Despite the existence of some quantitative differences in the Hubble functions of the two models, at least on a qualitative level the two descriptions give relatively similar results. The two approaches work well in the redshift range $z\in [0,0.4]$, where for some specific numerical values of the parameter $\lambda$ the Hubble function of the $f(Q,T)$ theory basically coincides with that of the $\Lambda$CDM model.   However, at larger redshifts significant differences in the cosmological evolution predicted by the two models do appear.

\begin{figure}
 \centering
 \includegraphics[scale=0.70]{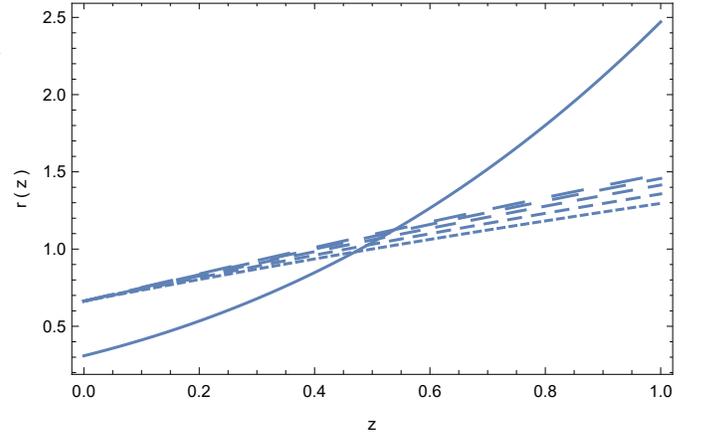}
 \caption{Variation of the matter energy density $\rho$  as a function of the redshift $z$ in the $f(Q,T)$ gravity theory for $f(Q,T)=-\alpha Q-\beta T^2$ and for a dust Universe,  for different values of the parameter
 $\lambda $: $\lambda =1.52$ (dotted curve), $\lambda   =1.64$ (short dashed curve),
 $\lambda =1.75$ (dashed curve), $\lambda =1.83$ (long dashed curve), and $\lambda =1.88$ (ultra long dashed curve), respectively. The variation of the matter energy density in the standard $\Lambda$CDM model is also represented as the solid curve.}
 \label{fig2}
\end{figure}

The variation of the matter energy density, given as function of the redshift by the relation $\rho (z)=\left[8\pi/\beta \left(1+32\pi\right)\right]\lp\sqrt{1+u(z)}-1\rp$ is represented in Fig.~\ref{fig2}, for  $8\pi/\beta \left(1+32\pi\right)=1.6$.

For all adopted numerical values of the parameter $\lambda$ the energy density is a monotonically increasing function of the redshift (a monotonically decreasing function of the cosmological time). The increase is almost linear, and for small redshifts it is almost independent on the numerical values of $\lambda$. However, a dependence on the model parameter can be seen at higher redshifts. The comparison with the matter energy density in the $\Lambda$CDM model shows that, if in the range $z\in [0,0.4]$ there is an approximate concordance between the predictions of the two models, for higher redshifts the differences in the matter densities are high. While in the $\Lambda$CDM model the matter energy density increases rapidly as $(1+z)^3$, the almost linear increase of $\rho$ in this particular $f(Q,T)$ model predicts a much lower matter density at higher redshifts.

\begin{figure}
 \centering
 \includegraphics[scale=0.70]{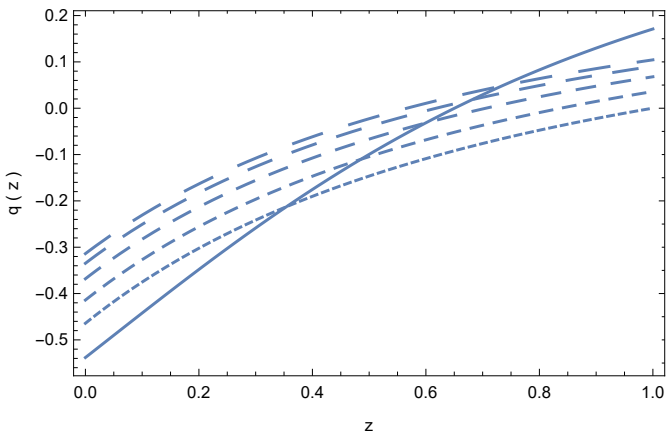}
 \caption{Variation of the deceleration parameter $q$ as a function of the redshift $z$ in the $f(Q,T)$ gravity theory for $f(Q,T)=-\alpha Q-\beta T^2$ and for a dust Universe,  for different values of the parameter
 $\lambda $: $\lambda =1.52$ (dotted curve), $\lambda   =1.64$ (short dashed curve),
 $\lambda =1.75$ (dashed curve), $\lambda =1.83$ (long dashed curve), and $\lambda =1.88$ (ultra long dashed curve), respectively. The variation of the deceleration parameter in the standard $\Lambda$CDM model is also represented as the solid curve.}
 \label{fig3}
\end{figure}

The variation of the deceleration parameter $q$ is represented, as a function of the redshift, in Fig.~\ref{fig3}, for the same values of the parameter $\lambda$ as considered in the previous figures.

The deceleration parameter is a monotonically increasing function of $z$. The evolution of the Universe begins, at redshift $z=1$, from a decelerating phase, with $q>0$. The expansion of the Universe accelerates, and at a finite value of $z$ it reaches the value $q=0$, corresponding to the transition to the accelerated phase. The evolution  of $q$  is strongly dependent on the numerical values of the model parameter $\lambda$. Depending on these values a large range of present day values of the deceleration parameter can be obtained. The comparison with the deceleration parameter variation in the $\Lambda$CDM model show that there is a qualitative similarity between the two models. However, the present choice of the function $f(Q,T)$ cannot fully reproduce the standard cosmological evolution, in a quantitative way. However, it provides similar qualitative results. Hopefully, by using an advanced fitting procedure, based on the direct application of observational data, and more general functional forms of $f(Q,T)$, this gravity theory may provide an alternative to the standard $\Lambda$CDM model.

\section{Discussions and final remarks}\label{sect4}

After more than one hundred years since Einstein did propose the first geometric theory of gravity, general theory, we are presently witnessing the very interesting situation that at least three geometric descriptions of gravity are possible, based on the three basic quantities introduced in Riemannian geometry and its extensions (curvature, torsion and nonmetricity, respectively). These findings raise the fundamental question of the possibility of a unique geometric description of gravity. Are these three descriptions completely equivalent, or they are perhaps just some particular cases of a more general geometric theory, which is still needed to be found?

In the present paper we have investigated some theoretical aspects of the third geometric description of gravity, the symmetric teleparallel gravity, or $f(Q)$ gravity, by  introducing a new class of theories where the nonmetricity $Q$ is coupled nonminimally to the trace of the matter energy-momentum tensor. From a mathematical point of view we have performed our analysis in the framework of the
metric-affine formalism. Our theory is constructed in a similar way like the $f(R,T)$ theory \cite{book}, but with the geometric part of the action being replaced by the symmetric teleparallel formulation. Similarly to the  in the standard curvature - trace of energy-momentum tensor couplings, in the $f(Q,T)$ theory the coupling  between $Q$ and $T$ leads to the nonconservation of the energy-momentum tensor. This nonconservation has important physical implications, implying significant changes in the thermodynamics of the Universe, similarly to those in the $f(R,T)$ theory \cite{book}, and, due to the nongeodesic motion of test particles,  to the appearance of an extra force. Our approach may also lead to an improvement of the
geometrical formulation of gravity theories with geometry-matter coupling. Implemented in both matter and geometry sectors, our approach allows a  consistent and workable representation of the nonminimal curvature-matter
coupling theories. In this context we have derived the gravitational field equations of the $f(Q,T)$ gravity theory from a variational principle that generalizes the variational principle of the $f(Q,T)$ theory, and we have obtained the general relation describing the nonconservation of the matter energy-momentum tensor.

As a theoretical test of our theory we have analysed its cosmological applications. As a first result in this respect we have obtained the
generalized Friedmann equations of the $f(Q,T)$ theory describing the cosmological evolution in a flat, homogeneous and isotropic  Friedmann-Lemaitre-Robertson-Walker type geometry. The generalized Friedmann equations can be reformulated as the standard equations of general relativity, but with the ordinary matter energy density and pressure replaced by some effective quantities. The effective quantities depend on the Lagrangian $f$ of the theory, and on its derivatives with respect to $Q$ and $T$. Interestingly enough, both effective thermodynamic quantities contain linear combinations of the ordinary matter energy density and pressure.  In fact the  coupling between the trace of the energy-momentum tensor  and the $Q$ field introduces two types of corrections. The first is the presence of an additive term of the form $f/4F$ that independently appears in both Friedmann equations. Secondly, we have the term $4\pi/F$, multiplying the linear combination constructed from the components of the
energy-momentum tensor (energy density and pressure) in both Friedmann equations. The coefficients of the linear combinations of the energy density and pressure are constructed from the function $\tilde{G}\propto f_T$. Consequently,  the basic equations describing
the cosmological evolution in $f(Q,T)$ gravity can be formulated in terms of an effective energy density and pressure, which
both depend on the energy and pressure components of the energy-momentum tensor, and on the functions $f(Q,T)$, $f_Q(Q,T)$ and $f_T(Q,T)$, respectively. An important indicator of the cosmological evolution, the deceleration parameter, given by Eq.~(\ref{58}), has a complicated dependence on the Lagrangian $f$ and its derivatives, indicating that, depending on the functional form of $f(Q,T)$, a large number of cosmological
 models can be obtained. We have also shown explicitly that for the vacuum case, when the $f(Q,T)$ theory reduces to $f(Q)$ theory, for late times, the Universe enters into an exponentially accelerating de Sitter type phase.

We have also considered three explicit classes of cosmological models, obtained by imposing some specific simple mathematical forms for the function $f(Q,T)$. In all three example we have considered that $Q$ and $T$ enter in an additive form in the Lagrangian, neglecting the possible existence of some cross terms of the type $QT$, or functions of it. In the case  $f(Q,T)=\alpha Q+\beta T$, the cosmological evolution is of de Sitter type, with the Universe expanding exponentially. The model $f(Q,T)=\alpha Q^n+\beta T$ leads to a power law type form of the scale factor, and to a constant deceleration parameter. However,  by an appropriate
choice of the model parameters $\alpha $, $\beta$ and $\gamma$ accelerating expansions can be
obtained easily. The third model with $f(Q,T)=-\alpha Q-beta T^2$ leads to a complex cosmological dynamics, involving the transition from a decelerating to an accelerating state. The results
can be obtained only by numerically integrating the generalized Friedmann equations. The nature of the cosmological evolution is strongly dependent on the numerical values of the model parameters. We have also compared the theoretical predictions of the $f(Q,T)$ theory with the corresponding results in the standard $\Lambda$CDM cosmology. For the
specific range of cosmological parameters we have considered it did follow that the Universe began its recent evolution in a decelerating phase,
and in the large time limit it can reach a de Sitter phase. Depending on the model parameters, a large spectrum of present day values of the deceleration parameter can be obtained. The theoretical predictions of the Hubble parameter are similar to those of the standard general relativistic cosmological model in the presence of the cosmological constant. However, some significant deviations appear for the behavior of the matter energy density and of the deceleration parameter. But if investigated for a larger range of parameters and functional forms of $f$ this model may represent an alternative to the $\Lambda$CDM cosmology, with the late time
de Sitter phase induced by the coupling between nonmetricity and matter.

The $f(Q,T)$  gravity theory is also valid when instead of ordinary matter one includes scalar fields in the action. Another possible application of the $f(Q,T)$ theory is to consider inflation in the presence of scalar fields, an approach that may provide a completely new perspective
on the geometrical, gravitational, and cosmological processes that did play a major role in the very early
dynamics  of the Universe.  The analysis of structure formation in $f(Q,T)$ theory is also a major topics of research that could be investigated, with the use of a  background metric. For different nonmetricity-trace of the energy-momentum tensor coupling models,
the SNIa, BAO, CMB shift parameter data can be used
to obtain constraints for the respective models. Moreover, such an approach may allow the detailed exploration and analysis of structure formation from a different perspective. An interesting issue is to obtain the Newtonian and
the post-Newtonian limits of the $f(Q,T)$ gravity, and to investigate the constraints the local
gravity at the Solar System level impose on the theory. The Newtonian limit can also help in finding constraints
arising from other astrophysical observations.

To conclude, in the present investigation we have introduced a new version of the symmetric teleparallel theory, and we have proven its theoretical consistency. This approach also motivates and encourages the study of further extensions of
the $f(Q)$ type family of theories. We have shown that the presented approach predicts de Sitter type expansions of the Universe, and thus it may represent a geometric alternatives to dark energy. Hence this study offers some basic theoretical tools for the in depth investigation of the geometric aspects of gravity, and of its cosmological implications.

\section*{Acknowledgments}

T. H. would like to thank the Yat-Sen School of the Sun Yat-Sen University
in Guangzhou, P. R. China, for the kind hospitality offered during the
preparation of this work. S.-D. L. thanks the Natural Science Foundation of Guangdong Province for financial support (grant No. 2016A030313313).

\appendix

\section{Calculation of $Q=-Q_{\alpha\mu\nu}P^{\alpha\mu\nu}$}\label{app1}
According to Eq.~(\ref{eq:Q} )and Eq.~(\ref{eq:disformation}), we have
\begin{eqnarray}
&&Q\equiv -g^{\mu \nu}\lp L^{\alpha}_{\ \ \beta\mu}L^{\beta}_{\ \ \nu\alpha} - L^{\alpha}_{\ \ \beta\alpha} L^{\beta}_{\ \ \mu \nu}  \rp , \\
&& L^{\alpha}_{\ \ \beta \mu}= -\frac{1}{2} g^{\alpha \lambda}\lp Q_{\mu\beta\lambda} +Q_{\beta\lambda\mu} - Q_{\lambda\mu\beta} \rp , \\
&& L^{\beta}_{\ \ \nu \alpha}= -\frac{1}{2} g^{\beta \rho}\lp Q_{\alpha\nu\rho} +Q_{\nu\rho\alpha} - Q_{\rho\alpha\nu} \rp , \\
&&  L^{\alpha}_{\ \ \beta \alpha}= -\frac{1}{2} g^{\alpha \lambda}\lp Q_{\alpha\beta\lambda} +Q_{\beta\lambda\alpha} - Q_{\lambda\alpha\beta} \rp \nonumber \\
&& = -\frac{1}{2}\lp \tilde{Q}_{\beta} + Q_{\beta} - \tilde{Q}_{\beta} \rp = -\frac{1}{2} Q_{\beta}, \\
&& L^{\beta}_{\ \ \mu\nu}= -\frac{1}{2} g^{\beta \rho}\lp Q_{\nu\mu\rho} +Q_{\mu\rho\nu} - Q_{\rho\nu\mu} \rp .
\end{eqnarray}
Thus, we obtain
\begin{eqnarray}
&&-g^{\mu \nu} L^{\alpha}_{\ \ \beta\mu}L^{\beta}_{\ \ \nu\alpha}= -\frac{1}{4}g^{\mu\nu}g^{\alpha\lambda}g^{\beta\rho}\lp Q_{\mu\beta\lambda} +Q_{\beta\lambda\mu} - Q_{\lambda\mu\beta} \rp \nonumber \\
&& \times \lp Q_{\alpha\nu\rho} +Q_{\nu\rho\alpha} - Q_{\rho\alpha\nu} \rp = -\frac{1}{4} \lp Q^{\nu\rho\alpha} +Q^{\rho\alpha\nu}-Q^{\alpha\nu\rho}  \rp \nonumber \\
&&\times \lp Q_{\alpha\nu\rho} +Q_{\nu\rho\alpha} - Q_{\rho\alpha\nu} \rp = -\frac{1}{4} (  \cancel{Q^{\nu\rho\alpha}Q_{\alpha\nu\rho}} +Q^{\nu\rho\alpha}Q_{\nu\rho\alpha} \nonumber \\
&& \bcancel{- Q^{\nu\rho\alpha}Q_{\rho\alpha\nu}}  + Q^{\rho\alpha\nu}Q_{\alpha\nu\rho} +\bcancel{Q^{\rho\alpha\nu}Q_{\nu\rho\alpha}} - Q^{\rho\alpha\nu}Q_{\rho\alpha\nu} \nonumber \\
&& -Q^{\alpha\nu\rho}Q_{\alpha\nu\rho} \cancel{-Q^{\alpha\nu\rho}Q_{\nu\rho\alpha}} +Q^{\alpha\nu\rho} Q_{\rho\alpha\nu})  \nonumber \\
&& =  -\frac{1}{4}\lp 2 Q^{\alpha\nu\rho} Q_{\rho\alpha\nu}-Q^{\alpha\nu\rho}Q_{\alpha\nu\rho} \rp , \\
&&g^{\mu \nu}L^{\alpha}_{\ \ \beta\alpha} L^{\beta}_{\ \ \mu \nu}= \frac{1}{4}g^{\mu\nu}g^{\beta\rho}Q_{\beta}\lp Q_{\nu\mu\rho} +Q_{\mu\rho\nu} - Q_{\rho\nu\mu} \rp \nonumber \\
&& = \frac{1}{4}Q^{\rho}\lp 2\tilde{Q}_{\rho}-Q_{\rho}  \rp, \\
&& Q =- \frac{1}{4}\lp -Q^{\alpha\nu\rho}Q_{\alpha\nu\rho}+ 2 Q^{\alpha\nu\rho} Q_{\rho\alpha\nu} - 2Q^{\rho}\tilde{Q}_{\rho} + Q^{\rho}Q_{\rho}  \rp. \nonumber \\
\label{eq:deltaQ}
\end{eqnarray}
Then, according to Eq.~(\ref{eq:superpotential}), we have
\begin{eqnarray}
&&P^{\alpha\mu\nu}=  \frac{1}{4}\bigg[  -Q^{\alpha \mu \nu}+ Q^{\mu \alpha  \nu} +Q^{\nu \alpha  \mu} + Q^{\alpha}g^{\mu \nu} - \tilde{Q}^{\alpha}g^{\mu\nu}\nonumber \\
&&
-\frac{1}{2}\lp  g^{\alpha\mu}Q^{\nu} + g^{\alpha\nu}Q^{\mu}  \rp \bigg ], \\
&& -Q_{\alpha\mu\nu}P^{\alpha\mu\nu}= -\frac{1}{4}\bigg[  -Q_{\alpha\mu\nu}Q^{\alpha \mu \nu}+ Q_{\alpha\mu\nu}Q^{\mu \alpha  \nu} \nonumber \\
&&+Q_{\alpha\mu\nu}Q^{\nu \alpha  \mu} + Q_{\alpha\mu\nu}Q^{\alpha}g^{\mu \nu} - Q_{\alpha\mu\nu}\tilde{Q}^{\alpha}g^{\mu\nu} \nonumber \\
&&-\frac{1}{2}Q_{\alpha\mu\nu}\lp  g^{\alpha\mu}Q^{\nu} + g^{\alpha\nu}Q^{\mu}  \rp \bigg ] = -\frac{1}{4}(-Q_{\alpha\mu\nu}Q^{\alpha \mu \nu}\nonumber \\
&&+2 Q_{\alpha\mu\nu}Q^{\mu \alpha  \nu} + Q_{\alpha}Q^{\alpha}-2Q_{\alpha}\tilde{Q}^{\alpha}) =Q.\label{eq:Qproof}
\end{eqnarray}

To obtain the above result we have used the relations  $Q_{\alpha\mu\nu}Q^{\mu \alpha  \nu}=Q_{\alpha\mu\nu}Q^{\nu \alpha  \mu}$, which is valid since $Q_{\alpha\mu\nu}Q^{\mu \alpha  \nu}=Q_{\alpha\nu\mu}Q^{\mu \alpha  \nu}=Q^{\alpha\nu\mu}Q_{\mu \alpha  \nu}=Q^{\nu\mu\alpha}Q_{\alpha  \nu\mu}=Q_{\alpha\mu\nu}Q^{\nu \alpha  \mu}$. Hence, we have proved that $Q=-Q_{\alpha\mu\nu}P^{\alpha\mu\nu}$, a relation which is very useful in later calculations.

\section{Calculation of the variation of $\delta Q$}\label{app2}

Before the presentation of the detailed variation of $\delta Q$, we write down all the nonmetricity tensors for later applications. They are obtained as
\begin{eqnarray}
\hspace{-0.8cm}&&Q_{\alpha\mu\nu}=\nabla_{\alpha}g_{\mu\nu}, \\
\hspace{-0.8cm}&&Q^{\alpha}_{\ \ \mu\nu}=g^{\alpha\beta}Q_{\beta\mu\nu}=g^{\alpha\beta}\nabla_{\beta} g_{\mu\nu}=\nabla^{\alpha}g_{\mu\nu}, \\
\hspace{-0.8cm}&& Q_{\alpha \ \ \nu}^{\ \ \mu}=g^{\mu\rho}Q_{\alpha\rho\nu}=g^{\mu\rho}\nabla_{\alpha}g_{\rho\nu} = - g_{\rho\nu} \nabla_{\alpha}g^{\mu\rho},\\
\hspace{-0.8cm}&& Q_{\alpha\mu}^{\ \ \ \ \nu}=g^{\nu\rho}Q_{\alpha\mu\rho}=g^{\nu\rho}\nabla_{\alpha}g_{\mu\rho}=-g_{\mu\rho}\nabla_{\alpha}g^{\nu\rho}, \\
\hspace{-0.8cm}&& Q^{\alpha\mu}_{\ \ \ \ \nu}=g^{\alpha\beta}g^{\mu\rho}\nabla_{\beta}g_{\rho\nu}=g^{\mu\rho}\nabla^{\alpha}g_{\rho\nu}=-g_{\rho\nu}\nabla^{\alpha}g^{\mu\rho},\nonumber \\ \\
\hspace{-0.8cm}&& Q^{\alpha \ \ \nu}_{\ \ \mu}= g^{\alpha\beta}g^{\nu\rho}\nabla_{\beta}g_{\mu\rho}=g^{\nu\rho}\nabla^{\alpha}g_{\mu\rho}=-g_{\mu\rho}\nabla^{\alpha}g^{\nu\rho} ,\nonumber \\ \\
\hspace{-0.8cm}&&Q_{\alpha}^{\ \ \mu\nu}=g^{\mu\rho}g^{\nu\sigma}\nabla_{\alpha}g_{\rho\sigma}=-g^{\mu\rho}g_{\rho\sigma}\nabla_{\alpha}g^{\nu\sigma}=-\nabla_{\alpha}g^{\mu\nu}, \nonumber \\ \\
\hspace{-0.8cm}&& Q^{\alpha\mu\nu}=-\nabla^{\alpha}g^{\mu\nu}
\end{eqnarray}

Let us find the variation of $Q$ by using Eq.~(\ref{eq:deltaQ}),
\begin{eqnarray}\label{eq:deltaQre}
&&\delta Q \nonumber \\
&&= - \frac{1}{4}\delta\lp -Q^{\alpha\nu\rho}Q_{\alpha\nu\rho}+ 2 Q^{\alpha\nu\rho} Q_{\rho\alpha\nu} - 2Q^{\rho}\tilde{Q}_{\rho} + Q^{\rho}Q_{\rho}  \rp \nonumber \\
&& =-\frac{1}{4}( -\delta Q^{\alpha\nu\rho}Q_{\alpha\nu\rho} -Q^{\alpha\nu\rho}\delta Q_{\alpha\nu\rho} + 2 \delta Q^{\alpha\nu\rho} Q_{\rho\alpha\nu} \nonumber \\
&& + 2 Q^{\alpha\nu\rho} \delta Q_{\rho\alpha\nu}- 2 \delta Q^{\rho}\tilde{Q}_{\rho}-2Q^{\rho}\delta \tilde{Q}_{\rho} + \delta Q^{\rho}Q_{\rho} \nonumber\\
&&+ Q^{\rho}\delta Q_{\rho})  \nonumber \\
&&= -\frac{1}{4}\Big[ Q_{\alpha\nu\rho}  \nabla^{\alpha}\delta g^{\nu\rho}-Q^{\alpha\nu\rho} \nabla_{\alpha}\delta g_{\nu\rho} - 2 Q_{\rho\alpha\nu} \nabla^{\alpha}\delta g^{\nu\rho} \nonumber \\
&&+ 2 Q^{\alpha\nu\rho} \nabla_{\rho}\delta g_{\alpha\nu}- 2\tilde{Q}_{\rho} \delta (-g_{\mu\nu}\nabla^{\rho}g^{\mu\nu})-2Q^{\rho}\delta (\nabla^{\lambda} g_{\rho\lambda}) \nonumber \\
&&+Q_{\rho} \delta (-g_{\mu\nu}\nabla^{\rho}g^{\mu\nu})+ Q^{\rho}\delta (-g_{\mu\nu}\nabla_{\rho}g^{\mu\nu})\Big] \nonumber \\
&&= -\frac{1}{4}\Big[ Q_{\alpha\nu\rho}  \nabla^{\alpha}\delta g^{\nu\rho}-Q^{\alpha\nu\rho} \nabla_{\alpha}\delta g_{\nu\rho} - 2 Q_{\rho\alpha\nu} \nabla^{\alpha}\delta g^{\nu\rho} \nonumber \\
&&+ 2 Q^{\alpha\nu\rho} \nabla_{\rho}\delta g_{\alpha\nu}+ 2\tilde{Q}_{\rho}\nabla^{\rho}g^{\mu\nu} \delta g_{\mu\nu}+ 2\tilde{Q}_{\rho}  g_{\mu\nu}\nabla^{\rho}\delta g^{\mu\nu}\nonumber \\
&&-2Q^{\rho} \nabla^{\lambda}\delta g_{\rho\lambda} -Q_{\rho}\nabla^{\rho}g^{\mu\nu} \delta g_{\mu\nu} -  Q_{\rho}  g_{\mu\nu}\nabla^{\rho}\delta g^{\mu\nu} \nonumber \\
&&- Q^{\rho}\nabla_{\rho}g^{\mu\nu}\delta g_{\mu\nu}   - Q^{\rho} g_{\mu\nu}\nabla_{\rho}\delta g^{\mu\nu} \Big] \nonumber \\
&&= -\frac{1}{4}\Big[ Q_{\alpha\nu\rho}  \nabla^{\alpha}\delta g^{\nu\rho}-Q^{\alpha\nu\rho} \nabla_{\alpha}\delta g_{\nu\rho} - 2 Q_{\rho\alpha\nu} \nabla^{\alpha}\delta g^{\nu\rho} \nonumber \\
&&+ 2 Q^{\alpha\nu\rho} \nabla_{\rho}\delta g_{\alpha\nu}+ 2\tilde{Q}_{\rho}\nabla^{\rho}g^{\mu\nu} \delta g_{\mu\nu}+ 2\tilde{Q}_{\rho}  g_{\mu\nu}\nabla^{\rho}\delta g^{\mu\nu}\nonumber \\
&&-2Q^{\rho} \nabla^{\lambda}\delta g_{\rho\lambda} -Q_{\rho}\nabla^{\rho}g^{\mu\nu} \delta g_{\mu\nu} -  Q_{\rho}  g_{\mu\nu}\nabla^{\rho}\delta g^{\mu\nu} \nonumber \\
&&- Q^{\rho}\nabla_{\rho}g^{\mu\nu}\delta g_{\mu\nu}   - Q^{\rho} g_{\mu\nu}\nabla_{\rho}\delta g^{\mu\nu} \Big].
\end{eqnarray}

In order to simplify the above equation we can use several useful equations, which are given below as
\begin{eqnarray}
&&\delta g_{\mu\nu}=-g_{\mu\alpha}\delta g^{\alpha\beta} g_{\beta\nu}, \\
&&-Q^{\alpha\nu\rho}\nabla_{\alpha}\delta g_{\nu\rho}=-Q^{\alpha\nu\rho}\nabla_{\alpha} \lp -g_{\nu\lambda}\delta g^{\lambda\theta} g_{\theta\rho}   \rp \nonumber \\
&& = 2 Q^{\alpha\nu}_{\ \ \ \ \theta}Q_{\alpha\nu\lambda} \delta g^{\lambda\theta}+Q_{\alpha\lambda\theta}\nabla^{\alpha}g^{\lambda\theta}\nonumber \\
&&=2 Q^{\alpha\sigma}_{\ \ \ \ \nu}Q_{\alpha\sigma\mu} \delta g^{\mu\nu}+Q_{\alpha\nu\rho}\nabla^{\alpha}g^{\nu\rho}, \\
&&  2 Q^{\alpha\nu\rho} \nabla_{\rho}\delta g_{\alpha\nu}=-4 Q_{\mu}^{\ \ \sigma\rho}Q_{\rho\sigma\nu}\delta g^{\mu\nu} -2 Q_{\nu\rho\alpha}\nabla^{\alpha}\delta g^{\nu\rho}, \nonumber \\ \\
&&-2Q^{\rho} \nabla^{\lambda}\delta g_{\rho\lambda}=2 Q^{\alpha}Q_{\nu\alpha\mu}\delta g^{\mu\nu} + 2 Q_{\mu}\tilde{Q}_{\nu}\delta g^{\mu\nu}  \nonumber \\
&&+ 2 Q_{\nu}g_{\alpha\rho}\nabla^{\alpha}g^{\nu\rho}.
\end{eqnarray}

Thus, Eq.~(\ref{eq:deltaQre}) takes the form
\begin{eqnarray}
&&\delta Q  \nonumber \\
&&= -\frac{1}{4}\Big[ Q_{\alpha\nu\rho}  \nabla^{\alpha}\delta g^{\nu\rho}+2 Q^{\alpha\sigma}_{\ \ \ \ \nu}Q_{\alpha\sigma\mu} \delta g^{\mu\nu}+Q_{\alpha\nu\rho}\nabla^{\alpha}g^{\nu\rho} \nonumber \\
&&- 2 Q_{\rho\alpha\nu} \nabla^{\alpha}\delta g^{\nu\rho} -4 Q_{\mu}^{\ \ \sigma\rho}Q_{\rho\sigma\nu}\delta g^{\mu\nu} -2 Q_{\nu\rho\alpha}\nabla^{\alpha}\delta g^{\nu\rho}  \nonumber \\
&&+ 2\tilde{Q}^{\rho}Q_{\rho\mu\nu} \delta g^{\mu\nu}+ 2\tilde{Q}_{\alpha}  g_{\nu\rho}\nabla^{\alpha}\delta g^{\nu\rho}+2 Q^{\alpha}Q_{\nu\alpha\mu}\delta g^{\mu\nu} \nonumber \\
&& + 2 Q_{\mu}\tilde{Q}_{\nu}\delta g^{\mu\nu} + 2 Q_{\nu}g_{\alpha\rho}\nabla^{\alpha}g^{\nu\rho}  -Q^{\rho}Q_{\rho\mu\nu} \delta g^{\mu\nu}  \nonumber \\
&&-  Q_{\alpha}  g_{\nu\rho}\nabla^{\alpha}\delta g^{\nu\rho}- Q^{\rho}Q_{\rho\mu\nu}\delta g^{\mu\nu}   - Q_{\alpha} g_{\nu\rho}\nabla^{\alpha}\delta g^{\nu\rho} \Big] \nonumber \\
&& = 2 P_{\alpha\nu\rho} \nabla^{\alpha}\delta g^{\nu\rho} - \lp P_{\mu\alpha\beta}Q_{\nu}^{\ \ \alpha\beta} -2 Q^{\alpha\beta}_{\ \ \ \ \mu}P_{\alpha\beta\nu} \rp \delta g^{\mu\nu}, \nonumber \\
\end{eqnarray}
where we have used the relations
\begin{eqnarray}
&&2 P_{\alpha\nu\rho}= -\frac{1}{4}\big [ 2 Q_{\alpha\nu\rho} -2 Q_{\rho\alpha\nu}-2 Q_{\nu\rho\alpha}  \nonumber \\
&&+ 2 (\tilde{Q}_{\alpha} - Q_{\alpha}) g_{\nu \rho}+ 2 Q_{\nu}g_{\alpha\rho}  \big ], \\
&&4\lp P_{\mu\alpha\beta}Q_{\nu}^{\ \ \alpha\beta} -2 Q^{\alpha\beta}_{\ \ \ \ \mu}P_{\alpha\beta\nu} \rp = 2 Q^{\alpha\beta}_{\ \ \ \ \nu}Q_{\alpha\beta\mu}\nonumber \\
&& -4 Q_{\mu}^{\ \ \alpha\beta}Q_{\beta\alpha\nu} +2 \tilde{Q}^{\alpha}Q_{\alpha\mu\nu} + 2 Q^{\alpha}Q_{\nu\alpha\mu} \nonumber \\
&&+2Q_{\mu}\tilde{Q}_{v}-Q^{\alpha}Q_{\alpha\mu\nu}.
\end{eqnarray}

\section{Variation of the gravitational action with respect to the connection}\label{app3}

The full action of the $f(Q,T)$ theory supplemented  with the Lagrangian multipliers is
\begin{eqnarray}
S= \int \md ^4 x \bigg[ \frac{ \sqrt{-g}}{16 \pi} f(Q, T) + \mathcal{L}_{M} \sqrt{-g}\nonumber \\
+ \lambda_{\alpha}^{\ \ \beta\gamma} T^{\alpha}_{\ \ \beta \gamma}  +\xi_{\alpha}^{\ \ \beta\mu\nu} R^{\alpha}_{\ \ \beta \mu\nu}  \bigg].
\end{eqnarray}
We can vary the action separately, thus obtaining
\begin{eqnarray}
&&\delta \bigg[  \frac{ \sqrt{-g}}{16 \pi} f(Q, T) + \mathcal{L}_{M} \sqrt{-g} \bigg]\nonumber \\
&&=  \bigg( \frac{4 \sqrt{-g}}{16 \pi} f_{Q} P^{\mu\nu}_{\ \ \ \ \alpha} +H_{\alpha}^{\ \ \mu \nu} \bigg)\delta \hat{\Gamma}^{\alpha}_{\ \ \mu\nu}\;, \\
&& \delta \big(\lambda_{\alpha}^{\ \ \mu\nu} T^{\alpha}_{\ \ \mu \nu} \big) = 2 \lambda_{\alpha}^{\ \ \mu\nu} \delta \hat{\Gamma}^{\alpha}_{\ \ \mu \nu}, \\
&& \delta \big( \xi_{\alpha}^{\ \ \beta\mu\nu} R^{\alpha}_{\ \ \beta \mu\nu} \big)=\xi_{\alpha}^{\ \ \beta\mu\nu} \Big[  \nabla_{\mu} \big( \delta \hat{\Gamma}^{\alpha}_{\ \ \nu\beta}\big)- \nabla_{\nu} \big( \delta \hat{\Gamma}^{\alpha}_{\ \ \mu\beta}\big)       \Big]\nonumber \\
&&=2\xi_{\alpha}^{\ \ \nu\beta\mu}\nabla_{\beta}\big(  \delta  \hat{\Gamma}^{\alpha}_{\ \ \mu\nu}   \big) \simeq  2\big( \nabla_{\beta} \xi_{\alpha}^{\ \ \nu\beta\mu}\big)  \delta  \hat{\Gamma}^{\alpha}_{\ \ \mu\nu} .
\end{eqnarray}

Thus,
\begin{eqnarray}
\delta S= \int \md ^4 x \bigg( \frac{4 \sqrt{-g}}{16 \pi} f_{Q} P^{\mu\nu}_{\ \ \ \ \alpha} +H_{\alpha}^{\ \ \mu \nu} +  2\lambda_{\alpha}^{\ \ \mu\nu} \nonumber \\
 + 2 \nabla_{\beta} \xi_{\alpha}^{\ \ \nu\beta\mu}    \bigg)    \delta  \hat{\Gamma}^{\alpha}_{\ \ \mu\nu} .
\end{eqnarray}

To eliminate the Lagrange multipliers, we take two covariant derivatives $\nabla_{\mu}\nabla_{\nu}$ or $\nabla_{\nu}\nabla_{\mu}$ (considering vanishing curvature tensor) of the integrand, and thus we finally arrive to Eq.~(\ref{eq:feqconnection}).

\section{Metric divergence of (1,1)-form field equations}\label{app4}

The metric divergence of the gravitational field equation Eq.~(\ref{eq:metricdivfeq}) of the $f(Q,T)$ theory is
\begin{eqnarray}
&&\D_{\mu}\Big[ f_{T}\big(T^{\mu}_{\ \ \nu}+\Theta^{\mu}_{\ \ \nu}\big)-8\pi T^{\mu}_{\ \ \nu}\Big]=\frac{1}{2}\p_{\nu}f \nonumber \\
&&+\D_{\mu}\Big(f_{Q}Q_{\nu}^{\ \ \alpha\beta}P^{\mu}_{\ \ \alpha\beta}\Big)+\D_{\mu}\bigg[\frac{2}{\g}\gra_{\alpha}\big(f_{Q}\g P^{\alpha\mu}_{\ \ \ \ \nu}\big) \bigg], \nonumber \\
\end{eqnarray}
where we have
\begin{eqnarray}
\hspace{-0.5cm}&&\D_{\mu}\Big(f_{Q}Q_{\nu}^{\ \ \alpha\beta}P^{\mu}_{\ \ \alpha\beta}\Big)=\nabla_{\mu}\Big(f_{Q}Q_{\nu}^{\ \ \alpha\beta}P^{\mu}_{\ \ \alpha\beta}\Big) \nonumber \\
\hspace{-0.5cm}&&+ \frac{1}{2}Q_{\mu}\Big(f_{Q}Q_{\nu}^{\ \ \alpha\beta}P^{\mu}_{\ \ \alpha\beta}\Big) + L^{\rho}_{\ \ \mu\nu}\Big(f_{Q}Q_{\rho}^{\ \ \alpha\beta}P^{\mu}_{\ \ \alpha\beta}\Big) ,\\
\hspace{-0.5cm}&& \D_{\mu}\bigg[\frac{2}{\g}\gra_{\alpha}\big(f_{Q}\g P^{\alpha\mu}_{\ \ \ \ \nu}\big) \bigg] \nonumber \\
\hspace{-0.5cm}&&=\frac{2}{\g}\D_{\mu}\bigg[\gra_{\alpha}\big(f_{Q}\g P^{\alpha\mu}_{\ \ \ \ \nu}\big) \bigg] \nonumber \\
\hspace{-0.5cm}&& = \frac{2}{\g} \nabla_{\mu}\gra_{\alpha}\big(f_{Q}\g P^{\alpha\mu}_{\ \ \ \ \nu}\big) \nonumber \\
\hspace{-0.5cm}&&+\frac{1}{\g} Q_{\mu}\gra_{\alpha}\big(f_{Q}\g P^{\alpha\mu}_{\ \ \ \ \nu}\big) \nonumber \\
\hspace{-0.5cm}&& +\frac{2}{\g} L^{\rho}_{\ \ \mu\nu}\gra_{\alpha}\big(f_{Q}\g P^{\alpha\mu}_{\ \ \ \ \rho}\big),
\end{eqnarray}
which gives
\begin{eqnarray}
&&\D_{\mu}\Big[ f_{T}\big(T^{\mu}_{\ \ \nu}+\Theta^{\mu}_{\ \ \nu}\big)-8\pi T^{\mu}_{\ \ \nu}\Big]+\frac{8\pi}{\g}\nabla_{\alpha}\nabla_{\mu}H_{\nu}^{\ \ \alpha\mu} \nonumber \\
&&=\frac{1}{2}\p_{\nu}f+\nabla_{\mu}\Big(f_{Q}Q_{\nu}^{\ \ \alpha\beta}P^{\mu}_{\ \ \alpha\beta}\Big)+\frac{1}{2}Q_{\mu}\Big(f_{Q}Q_{\nu}^{\ \ \alpha\beta}P^{\mu}_{\ \ \alpha\beta}\Big) \nonumber \\
&&+ L^{\rho}_{\ \ \mu\nu}\Big(f_{Q}Q_{\rho}^{\ \ \alpha\beta}P^{\mu}_{\ \ \alpha\beta}\Big) +\frac{2}{\g} L^{\rho}_{\ \ \mu\nu}\gra_{\alpha}\Big(f_{Q}\g P^{\alpha\mu}_{\ \ \ \ \rho}\Big)\nonumber \\
&& +\frac{1}{\g} Q_{\mu}\gra_{\alpha}\Big(f_{Q}\g P^{\alpha\mu}_{\ \ \ \ \nu}\Big)=\sum_{i=1}^{10} E_i.
\end{eqnarray}

For the sake of clarity, in the above equation we have defined
\begin{eqnarray}
&&E_1=\frac{1}{2}\p_{\nu}f ,\\
&&E_2=\Big(\nabla_{\mu}f_{Q}\Big)Q_{\nu\alpha\beta}P^{\mu\alpha\beta},\\
&&E_3= f_{Q}\Big( \nabla_{\mu}Q_{\nu\alpha\beta}\Big)P^{\mu\alpha\beta}, \\
&& E_4=f_{Q}Q_{\nu\alpha\beta}\Big(\nabla_{\mu}P^{\mu\alpha\beta}\Big),\\
&& E_5=\frac{1}{2}f_{Q} Q_{\mu}Q_{\nu\alpha\beta}P^{\mu\alpha\beta}, \\
&& E_6=f_{Q} L^{\rho}_{\ \ \mu\nu}Q_{\rho\alpha\beta}P^{\mu\alpha\beta},\\
&& E_7= 2\Big(\nabla_{\alpha}f_{Q}\Big) L^{\rho}_{\ \ \mu\nu}P^{\alpha\mu}_{\ \ \ \ \rho},\\
&& E_8=f_{Q}Q_{\alpha}L^{\rho}_{\ \ \mu\nu}P^{\alpha\mu}_{\ \ \ \ \rho}, \\
&& E_9=2 f_{Q}L^{\rho}_{\ \ \mu\nu}\nabla_{\alpha}P^{\alpha\mu}_{\ \ \ \ \rho},\\
&& E_{10}=\frac{1}{\g} Q_{\mu}\gra_{\alpha}\Big(f_{Q}\g P^{\alpha\mu}_{\ \ \ \ \nu}\Big).
\end{eqnarray}

Then, we can find the following relations
\begin{eqnarray}
&&E_2+E_7=\nabla_{\mu}f_{Q}\Big( Q_{\nu\alpha\beta} + 2 L_{\beta\alpha\nu}\Big)P^{\mu\alpha\beta}=0, \\
&&E_5+E_8=\frac{1}{2}f_{Q}Q_{\mu}\Big( Q_{\nu\alpha\beta} + 2 L_{\beta\alpha\nu}\Big)P^{\mu\alpha\beta}=0, \\
&&E_4+E_9=f_{Q}\bigg[Q_{\nu\alpha\beta}\Big(\nabla_{\mu}P^{\mu\alpha\beta}\Big) +2L^{\rho}_{\ \ \mu\nu}\nabla_{\alpha}P^{\alpha\mu}_{\ \ \ \ \rho}\bigg] \nonumber \\
&&=f_{Q}\bigg[\Big(Q_{\nu\alpha\beta} +2 L_{\beta\alpha\nu}  \Big)\nabla_{\mu}P^{\mu\alpha\beta} +2L^{\rho}_{\ \ \alpha\nu}Q_{\mu\beta\rho}P^{\mu\alpha\beta}\bigg]\nonumber \\
&&=2f_{Q}L^{\rho}_{\ \ \alpha\nu}Q_{\mu\beta\rho}P^{\mu\alpha\beta}, \\
&& E_3+E_6+E_4+E_9\nonumber \\
&&=f_{Q}\Big( \nabla_{\mu}Q_{\nu\alpha\beta}+2L^{\rho}_{\ \ \alpha\nu}Q_{\mu\beta\rho}+L^{\rho}_{\ \ \mu\nu}Q_{\rho\alpha\beta}\Big)P^{\mu\alpha\beta} \nonumber \\
&&=\frac{1}{2}f_{Q}\D_{\nu}\Big( Q_{\mu\alpha\beta} P^{\mu\alpha\beta}      \Big)=-\frac{1}{2}f_{Q}\p_{\nu}Q.
\end{eqnarray}

Finally, we obtain
\begin{eqnarray}
\D_{\mu}\Big[ f_{T}\big(T^{\mu}_{\ \ \nu}+\Theta^{\mu}_{\ \ \nu}\big)-8\pi T^{\mu}_{\ \ \nu}\Big]+\frac{8\pi}{\g}\nabla_{\alpha}\nabla_{\mu}H_{\nu}^{\ \ \alpha\mu}\nonumber \\
= \frac{1}{2}\p_{\nu}f-\frac{1}{2}f_{Q}\p_{\nu}Q + \frac{1}{\g} Q_{\mu}\gra_{\alpha}\Big(f_{Q}\g P^{\alpha\mu}_{\ \ \ \ \nu}\Big)\nonumber \\
= \frac{1}{2}f_{T}\p_{\nu}T+ \frac{1}{\g} Q_{\mu}\gra_{\alpha}\Big(f_{Q}\g P^{\alpha\mu}_{\ \ \ \ \nu}\Big).\nonumber \\
\end{eqnarray}

\section{Calculation of $ Q=6H^2/N^2$} \label{app5}

Recalling Eq.~(\ref{eq:Qproof}), we have
\begin{eqnarray}
Q=-\frac{1}{4}\Big(-Q_{\alpha\mu\nu}Q^{\alpha \mu \nu}+2 Q_{\alpha\mu\nu}Q^{\mu \alpha  \nu}\nonumber \\
 + Q_{\alpha}Q^{\alpha}-2Q_{\alpha}\tilde{Q}^{\alpha}\Big) .
\end{eqnarray}

By using the relations already presented  in Appendix~\ref{app2}, for the case of the Friedmann-Robertson-Walker metric we obtain
\begin{eqnarray}
&&-Q_{\alpha\mu\nu}Q^{\alpha \mu \nu}=\gra_{\alpha}g_{\mu\nu}\gra^{\alpha}g^{\mu\nu}=\frac{4}{N^2}\lp T^2 +3H^2\rp, \\
&&Q_{\alpha\mu\nu}Q^{\mu \alpha  \nu}=-\gra_{\alpha}g_{\mu\nu}\gra^{\mu}g^{\alpha\nu}=-\frac{4}{N^2}T^2, \\
&&Q_{\alpha}Q^{\alpha}=\lp g_{\rho\mu}\gra_{\alpha}g^{\rho\mu} \rp \lp g_{\sigma\nu}\gra^{\alpha}g^{\sigma\nu} \rp=-\frac{4}{N^2}\lp T+3H \rp^2, \nonumber\\
\\
&&Q_{\alpha}\tilde{Q}^{\alpha}=\lp g_{\mu\rho}\gra_{\alpha}g^{\mu\rho} \rp \lp \gra_{\beta}g^{\alpha\beta} \rp = -\frac{4}{N^2}\lp T^2 +3HT \rp . \nonumber \\
\end{eqnarray}

Thus, we have
\begin{eqnarray}
Q=-\frac{1}{4}\bigg[\frac{4}{N^2}\lp T^2 +3H^2\rp -\frac{4}{N^2}2T^2 \nonumber \\
-\frac{4}{N^2}\lp T+3H \rp^2+\frac{4}{N^2}\lp 2T^2 +6HT \rp \bigg]=6\frac{H^2}{N^2}.
\end{eqnarray}

\end{document}